\documentclass[journal,twocolumn]{IEEEtran}
\IEEEoverridecommandlockouts
\usepackage{cite}
\usepackage{url}
\usepackage{amsmath,amssymb,amsfonts,bm}
\usepackage{algorithmic}
\usepackage{graphicx}
\usepackage{textcomp}
\usepackage{xcolor}
\usepackage{tabularx}
\usepackage{graphicx}
\usepackage[caption=false,font=footnotesize]{subfig}  % IEEE-safe
\usepackage[acronyms,nomain]{glossaries}
\makeglossaries
\newacronym{TX}{TX}{transmitter}
\newacronym{RX}{RX}{receiver}
\newacronym{LOS}{LOS}{line-of-sight}
\newacronym{NLOS}{NLOS}{non-line-of-sight}
\newacronym{UE}{UE}{user equipment}
\newacronym{BS}{BS}{base station}
\newacronym{US}{US}{United States}
\newacronym{UK}{UK}{United Kingdom}
\newacronym{SIR}{SIR}{signal-to-interference ratio}
\newacronym{5G}{5G}{fifth generation}
\newacronym{UAV}{UAV}{unmanned aerial vehicle}
\newacronym{DEM}{DEM}{digital elevation model}
\newacronym{USGS}{USGS}{United States Geological Survey}
\newacronym{OSM}{OSM}{Open Street Maps}
\newacronym{MS}{MS}{Microsoft}
\newacronym{NLCD}{NLCD}{national land cover data}
\newacronym{SSE}{SSE}{sum of squared errors}
\newacronym{NSSE}{NSSE}{normalized sum of squared errors}
\newacronym{UMa}{UMa}{urban macrocell}
\newacronym{MetMa}{MetMa}{metropolitan macrocell}
\newacronym{SMa}{SMa}{suburban macrocell}
\newacronym{RMa}{RMa}{rural macrocell}
\newacronym{AIC}{AIC}{Akaike Information Criteria}
\newacronym{AICc}{AICc}{Akaike Information Criteria corrected}
\newacronym{ML}{ML}{maximum likelihood}
\newacronym{CDF}{CDF}{cumulative distribution function}
\newacronym{LTE}{LTE}{Long-Term Evolution}
\newacronym{NYU}{NYU}{New York University}
\newacronym{3D}{3D}{three dimensional}
\newacronym{GIS}{GIS}{geographic information system}
\newacronym{FCC}{FCC}{Federal Communications Commission}
\newacronym{ISD}{ISD}{inter-site distance}
\newacronym{MSE}{MSE}{mean square error}
\newacronym{MSLE}{MSLE}{mean square log error}
\newacronym{WMSE}{WMSE}{weighted MSE}
\usepackage[margin=0.72in,bottom=0.95in]{geometry}
\usepackage{pgfplots}
\pgfplotsset{compat=1.18}
\usepackage{tikz}
\usetikzlibrary{shapes.geometric, arrows, shapes.multipart}
\usetikzlibrary{external}
\DeclareMathOperator*{\argmax}{arg\,max}

\tikzstyle{startstop} = [rectangle, rounded corners, minimum width=3cm, minimum height=1cm,text centered, draw=black, fill=red!30]
\tikzstyle{io} = [trapezium, trapezium left angle=70, trapezium right angle=110, minimum width=3cm, minimum height=1cm, text centered, draw=black, fill=blue!30]
\tikzstyle{process} = [rectangle, minimum width=3cm, minimum height=1cm, text centered, draw=black, fill=orange!30]
\tikzstyle{decision} = [diamond, minimum width=3cm, minimum height=1cm, text centered, draw=black, fill=green!30]
\tikzstyle{arrow} = [thick,->,>=stealth]

\def\BibTeX{{\rm B\kern-.05em{\sc i\kern-.025em b}\kern-.08em
    T\kern-.1667em\lower.7ex\hbox{E}\kern-.125emX}}

\begin{document}

%%%%%%%%%%%%% OLD TITLE %%%%%%%%%%%%%%
% \title{A Statistical-Geospatial Framework for Large Scale Line-of-Sight Probability Modeling from Real Macrocell Deployments\\
% % {\footnotesize \textsuperscript{*}Note: Sub-titles are not captured in Xplore and
% % should not be used}
% % \thanks{Identify applicable funding agency here. If none, delete this.}
% }
%%%%%%%%% NEW TITLE %%%%%%%%%%%%%%%%
\title{Line-of-Sight Probability in Macrocells: Framework, Statistical Models, and Parametrization from Massive Real World Datasets in the USA\\
% {\footnotesize \textsuperscript{*}Note: Sub-titles are not captured in Xplore and
% should not be used}
% \thanks{Identify applicable funding agency here. If none, delete this.}
}

\author{Bassel Abou Ali Modad~\IEEEmembership{Student Member,~IEEE,}, Xin Yu, Yao-Yi Chiang, Andreas F. Molisch~\IEEEmembership{Fellow,~IEEE},
\thanks{B. Abou Ali Modad, X. Yu and A.F. Molisch are with the University of Southern California, Los Angeles, CA, USA (e-mail: aboualim@usc.edu, xyu059@usc.edu, molisch@usc.edu.
Y. Chiang is with the University of Minnesota Twin Cities, Mineapolis, MN, USA (e-mail: yaoyi@umn.edu).}
\thanks{The work in this paper was supported by the National Science Foundation. Part of this work was presented at IEEE Globecom 2023 \cite{globecom-los}.}
}

\maketitle

% \textcolor{red}{(Total: around 24.5 columns without references --$>$ around 26 with references (will be close to 30): around 13 pages total)}
\begin{abstract}
    Accurate modeling of \gls{LOS} probability is crucial for wireless channel description and coverage planning. The presence of a \gls{LOS} impacts other channel characteristics such as pathloss, fading depth, delay- and angular spread, etc.. Existing models, although useful, are based on very limited datasets. In this paper, we establish a framework to produce high accuracy \gls{LOS} models from geospatial data in different environments, and apply it to create a \gls{LOS} model for macrocells, using datasets of the \gls{US} on a {\em national} scale, using more than $13,000$ locations of real-world macrocells. Based on this we create a new, fully parameterized model that better describes macrocell deployments in the \gls{US} than the 3GPP model. We furthermore demonstrate that for improved accuracy the \gls{LOS} probability should be modeled on a {\em per cell} basis, and the model parameters treated as random variables; we provide a full description and parameterization of this novel approach and by simulations show that it better predicts the inter-cell interference at the cell-edge than an average-based model. 
\end{abstract}

\begin{IEEEkeywords}
Line-of-sight probability, channel models, 3GPP, geospatial data, deployments
\end{IEEEkeywords}

\section{Introduction} \label{sec:intro}
     % \textcolor{red}{((around 3.5 to 4 columns))}     
        % \item Importance of LOS for channel characteristics\\
       % 0 \textcolor{blue}{Existence of a line-of-sight (LOS) connection between the user equipment (UE) and the base station (BS) is important channel parameter. LOS channels having less pathloss, less powerful fading and lower delay dispersion \cite{molisch2011wireless}. LOS existence beneficial for TOA ranging \cite{Zekavat2011}.} \\
         Among the parameters required to accurately characterize wireless channels, one of the most important is the existence of a line-of-sight (LOS) connection between the \gls{UE} and the \gls{BS}. \gls{LOS} channels have lower pathloss, smaller fading depth and lower delay- and angular dispersion \cite{molisch2023wireless}. The presence of a \gls{LOS} is also a significant factor for the accuracy of various ranging methods, in particular those based on time-of-arrival or direction-of-arrival measurements \cite{Zekavat2011}. The effects of a \gls{LOS} connection become more pronounced at higher frequencies, such as used for  \gls{5G} and beyond \gls{5G} systems \cite{Rappaport2022}.
        
        % \item LOS and NLOS conditions especially important for the wireless channel at higher frequencies: 5G , THz \textcolor{blue}{(0.5 columns)}\\
        % \textcolor{blue}{
        % LOS/NLOS distinction more important at higher frequencies (deeper shadows in NLOS) e.g., FR3 . At all frequencies, impact particularly on cell-edge users for coverage (getting less power than necessary). Also  because of inter-cell interference (more interference power than anticipated). Mis-estimation of LOS probability leads to wrong system planning. Thus, LOS probability important for system simulation in standards and companies. Several models exist, each with pros and cons. 
        % } \\
        A particularly important effect is the coverage and cell-edge performance of both noise-limited and interference limited systems: for noise-limited systems, a higher percentage of locations with \gls{LOS} increases coverage, while in the interference-limited case, the reverse might be true: \gls{LOS} increases the inter-cell interference, leading to more outage and requiring a change of the network planning and \gls{BS} locations to mitigate. For system development, models for the \gls{LOS} might influence decisions such as length of cyclic prefix in OFDM systems, beamformer algorithms, etc., since the existence of \gls{LOS} impacts delay and angular dispersion, as mentioned above. 
        Therefore, system-level simulations require an accurate LOS probability model that reflects the real world conditions. \\
        \textcolor{black}{\gls{LOS} models (and channel models in general) can be either site-specific or statistical. Site-specific models will, by design, always give more precise information about the behavior of a particular cell, or at a specific location within a particular cell, and are thus the method of choice for, e.g., cell planning and analyzing problems in a particular location. This extends to aspects like digital twins that emulate the complete system behavior at that particular location. Statistical (stochastic) models, such as the one we present in this paper, are used to obtain essential insights about the system behavior in a much faster way. In some cases, one might be more interested in the probability that a certain channel parameter attains some value without worrying where exactly this value is attained, e.g., outage probability. Such a study is important for standardization since the design, testing and comparison of wireless systems needs models for the properties of the channel that affect system performance. Thus, for \textit{designing} (not deploying) of new wireless systems, statistical models are indispensable.
        Moreover, site-specific evaluations provide the input dataset used for the subsequent statistical modeling. This data is then used to obtain statistical information that can be represented in a simple way.}
        Several statistical models exist in the standards and literature, each with their own assumptions, advantages and drawbacks as we will outline next.
        
        %conditions drastically change in higher frequency channels: 
    \subsection{Literature review} 
    % \textcolor{blue}{(1.5 columns)}    \\
     This subsection presents an overview of the literature on \gls{LOS} probability modeling and standard models, along with the assumptions, methods and datasets used in deriving them.
    
        % \item Measurement-based models: \textcolor{blue}{(0.5 columns)}\\           %        
       % \item 3GPP model \cite{38901} origins: $d_1/d_2$ UMa model, later included UMi and RMa models \\
       % \textcolor{blue}{Original  3GPP specifications: defined a 2-D distance dependent model for urban macrocell environments based on a few European urban sites with low building- and BS height. 3GPP later added a rural model and a dependency on UE height based on ray tracing techniques in a simulated urban environment \cite{Townend-journal}. Note: 3GPP offers a microcell model as well, but not focus of our work (macrocell only)} \\
   \subsubsection{Measurement-based models}
    The first class of models are derived from actual measurements. The original 3GPP specifications defined a two-dimensional distance dependent model for urban macrocell environments based on a few European urban sites, which have relatively low building- and BS- height. The model is defined by two parameters $d_1$, the cutoff distance where the \gls{LOS} probability is 1, and $d_2$ the decay rate of the LOS probability after the distance $d_1$. 3GPP later added a rural model and a dependency on UE height based on ray tracing techniques in a simulated urban environment \cite{Townend-journal}. 3GPP offers a microcell model as well, but it is not our focus in this work, since we will deal with macrocell BSs only.
   % \item NYU squared model \cite{Rappaport,sun-rappaport} \\
   % \textcolor{blue}{Enhancement of 3GPP model with a squared term, based on measurements in dense urban Manhattan environments, however with the same BS height assumptions as 3GPP. Also adopted in 3GPP as  optional mode} \\
    The \gls{NYU} squared model \cite{Rappaport,sun-rappaport} is an extension to the 3GPP model with a squared term, which provides a better fit in some situations. It is based on measurements in dense urban Manhattan environments, however with the same \gls{BS} height assumptions as 3GPP. It is also adopted in 3GPP as an optional mode.
   % \item Difference between them: Square term provides better fit in some environments \\
   % \textcolor{blue}{Addressed above}
   % \item Drawbacks: Based on small number of points, limited number of BS heights due to assumptions of typical European cities where the building heights do not generally exceed 4 storeys, and the assumption of 25m BS height (still higher than the  buildings). Several measurements show larger  LOS probability at intermediate (not closer) distances not captured by the model, see below  \\
   % \textcolor{blue}{These models although useful and widely used, are based on a limited number of measurements in specific environments. BS heights do not match real deployments in either US or Asia. }\\
    These models, although useful and widely adopted, are based on a limited number of measurements in specific environments.  Additionally, the 3GPP model was based on typical European cities where the buildings are generally no taller than 4 storeys, while building structures and BS deployments for cities in the US and Asia (where buildings are considerably taller) were ignored.

    % \item Papers using ray tracing with Geospatial data/Point cloud data with ray tracing \textcolor{blue}{(0.5 columns)} \\
    \subsubsection{Ray tracing-based models with geospatial/point cloud data}
     Other works use ray tracing along with real data gathered from point clouds or geospatial surveys of buildings.\footnote{Note that the accuracy of ray tracing for \gls{LOS} existence is better than that of ray tracing for, e.g., field strength prediction because the problem is much simpler and depends on fewer parameters (e.g., material of obstacles does not play a role in optical \gls{LOS} computation).} While this approach can analyze a large number of datasets in various environments, the validity and quality of the results depend on the quality of the underlying datasets. 
     %%%%%%%%%%%%% New stuff Reviewer 2 %%%%%%%%%%%%%%%
     Note that in this category of models, some approaches use the actual BS locations in the ray tracing simulations, and some other approaches use fictitious locations as representative of real BS locations (e.g. lamp posts as will be discussed).
        % \begin{itemize}
            % \item Korea paper (UMi case) \cite{koreaUMi}: BS height of 10m, double exponential model, distances only up to 400m considered, bin size is too large 25m\\             
               
            % \item Rappaport: \cite{Rappaport} ray tracing simulation based on point cloud. Only consider distances of up to 200m, they state that other cities will have different parameters. BS heights went up to 17m only, rather simplistic environment (based on the provided figures in the paper) \\
            In the category of approaches that use real BS locations with ray tracing is the work on the \gls{NYU} model \cite{Rappaport}. The authors use ray tracing based on a \gls{3D} database of buildings in New York City in the environment to derive the \gls{LOS} probabilities over a 200m-radius area.
        
            % \item Haneda LOS based on point clouds \cite{hanedaLOSPcd}: considered multiple scenarios gathered from laser scanning, but did not consider large distances, used lamp-posts as BS \\
            Ref. \cite{hanedaLOSPcd} evaluated the \gls{LOS} probability in both indoor and outdoor scenarios using point cloud scans of the environments and lamp posts as \gls{BS} locations, and proposed a generic exponential model inspired by 3GPP.
            % \item Work in \cite{lithuania} proposed a dual-environment LOS model where the assumption is that environments are composed of two different areas with different building densities  and heights, the latter are modeled as normal distributions.\\
            Ref. \cite{lithuania} proposed a mixed-environment (environments with different building densities and heights) urban model based on geospatial data in San Francisco.
            Besides these works, some of the work in this category studies LOS probability models in urban microcell environments such as \cite{koreaUMi}, but uses a different model than 3GPP (sum of two exponentials).
            
            % \item Drawbacks: low BS heights due to being UMi \cite{koreaUMi, hanedaLOSPcd}, only considered small distances \cite{koreaUMi, hanedaLOSPcd, Rappaport}, or small heights with a rather simplistic environment \cite{Rappaport}. \\
            % \textcolor{blue}{Only low BS heights due to microcell environments (\cite{koreaUMi, hanedaLOSPcd}); short distances and low macro-BS height (17m) \cite{Rappaport} (authors state that the derived LOS parameters may be different in other environments). Furthermore, the work in \cite{lithuania} did not consider the LOS on the street level, but instead used areas with direct optical visibility to the BS as part of the LOS data.} \\
            However, those works do not reflect the fact that in urban macrocells, both BS height and building height are often large. The assumption of low BS height might be due to the fact that a microcellular environment is analyzed (\cite{koreaUMi, hanedaLOSPcd}), or be motivated by compatibility with the 3GPP model (e.g., in  \cite{Rappaport} the \gls{BS} height is 17m; the authors state that the derived LOS parameters may be different in other environments). Furthermore, the work in \cite{lithuania} did not consider the LOS on streets only, but instead included open areas with no streets or buildings in the derivation of their LOS model.
        % \end{itemize} 
    % \item ray tracing with geospatial data and fictitious locations: \\
    % \textcolor{blue}{Geospatial with fictitious BS locations, e.g., lampposts. Useful to see how particular BS deployment would work, but by definition not in agreement with real deployment.} \\
    % \subsubsection{Ray tracing-based models with geospatial data and fictitious \gls{BS} locations}
    
    Similarly to the above models, there are investigations using geospatial data, but with fictitious BS locations (such as lamp posts) to derive LOS models. This category of works is useful to see how particular BS deployments would work, but by definition, is not in agreement with real deployments.
        % \begin{itemize}
            % \item Work in Townend \cite{Townend-conf,Townend-journal, Townend3}: UK based geospatial data collected with Lidar, use raytracing and derive models for urban, suburban, rural areas (\cite{Townend-journal}) \\
            An important example for this type of investigation is a series of papers \cite{Townend-conf,Townend-journal, Townend3}, where geospatial data was collected from LiDAR surveys in the \GLS{UK} of multiple environment categories, namely urban, suburban, and rural.
            % \item They set up a framework for classification of cells into environmental categories based on ITU-R recommendations of inter-site distance (\cite{Townend-conf,Townend-journal}) analyzed some characteristics of each category such as building coverage distribution, neighbor count, antenna height, and derived an "average" model for each environment class. \\
            The authors developed a framework to classify cells into the above-mentioned categories based on ITU-R recommendations of \gls{ISD}, derived LOS models and analyzed building and deployment related characteristics in each environment.
            % \item They found very different results, and that 3GPP is not a good fit even when using MSE-fit for $d_1$, $d_2$ parameters \\
            Results showed that the 3GPP model with its default parameters ($d_1=18$ and $d_2=63$) was not a good fit, and that fitting the $d_1$ and $d_2$ parameters to the data leads to a better fit than the preset parameters. However, even the revised fit shows significant deviations from the raw data.
            % \item Drawbacks: Lamp posts were used as representative BS locations (not realistic). Also, results have limited applicability to US cities since European cities are generally denser and height distribution is different. 
            % Most importantly, Inter-site distance does not capture radio propagation characteristics of each environment category and is provider dependent, Thus not good measure for environment category. \\
            
            Although this work provides important contributions, it has some drawbacks. Most importantly, the chosen BS locations are not based on real deployments. In any case, the model is not representative of macrocell \gls{BS}s, since those are mounted on building rooftops at much greater height in practice, not on lampposts. Secondly, 
            the results have limited applicability to \gls{US} cities since the urban planning in the \gls{UK} is generally different in terms of building heights and coverage (European cities are generally denser). Finally, an environment classification based on \gls{ISD} does not capture radio propagation characteristics since it is an operator-dependent choice.

            \textcolor{black}{In this same category of models, there exist radio mapping tools such as Sionna \cite{sionna}, WinProp \cite{winprop}, SCADAcore \cite{scadacore}, and others that perform site-specific propagation prediction that could be used to generate \gls{LOS} labels in  a particular geographic area. While these are powerful tools that can help with radio propagation prediction and network planning, this is not the aim of our work, and using such tools to check \gls{LOS} status of street points would be equivalent to the method we use in the following section, since both our method and these tools use the same underlying geographical databases.}
        % \end{itemize}
    % \item Papers using Raytracing with Simulation data: \\
    % \textcolor{blue}{Alternatively, some works resort to simulated city environment to perform raytracing and study the LOS probability.  Faster than gathering and processing geospatial data or point clouds, but by definition not as realistic, and depends on assumptions. Used for both macrocells and UAV prediction} \\
    \subsubsection{Ray tracing-based models with simulation data}
    Alternatively, some works resort to simulated city environment to perform ray tracing and study the \gls{LOS} probability, mostly for \gls{UAV} communications. This method is faster than gathering and processing geospatial data or point clouds, but has some limitations.
    % \begin{itemize}
        % \item Geometry-based UAV LOS \cite{geometryUAV-alhourani}: Ray tracing simulation of urban environments for UAV communications at different heights.  The city environment is generated according to specified ITU-R parameters \cite{itu-r-rec} that define building coverage, heights and density statistics \\
        Ref. \cite{UAV-LOS-Sim} studies the \gls{LOS} probability in \gls{UAV} communications scenarios using simulated city data and ray tracing. The city environment is generated according to specified ITU-R parameters \cite{itu-r-rec} that define building coverage, heights and density statistics.
        % \item Work in \cite{UAV-LOS-Sim} uses the same parameters from \cite{itu-r-rec} to simulate an urban environment for UAV LOS. Derives LOS models for UAV communications at different heights and user-UAV distances. \\
        % \textcolor{red}{Similarly, \cite{UAV-LOS-Sim} proposes a new simulator based on the same building simulation parameters to derive LOS models for UAV communications at different heights and user-UAV distances.}
        % \item Drawbacks: recommendations in \cite{itu-r-rec} state their limitations: (i) no terrain variation taken into account, (ii) real density and coverage statistics greatly vary and do not always fit the suggested distributions, (iii) model distributions only hold for smaller areas, uniform building sizes (in larger cells we can have in reality  mixture of high-rise and lower height buildings). \\
        The ITU-R states the limitations of its simulation models, namely, the effect of terrain elevation is not taken into account, the building density and coverage statistics vary greatly and do not always fit the suggested distributions in \cite{itu-r-rec}, which only hold for regions with smaller areas according to \cite{itu-r-rec}. Moreover, we find that building sizes are not always uniform in urban areas in the \gls{US}, and in a large enough cell, we can have a mixture of high-rise and lower height buildings. Some other works such as \cite{los-grid} assume deterministic grid layouts and constant building heights and sizes in deriving their \gls{LOS} probability model which was validated against a few site-specific \gls{GIS} data.
    % \end{itemize}
    % \item Stochastic geometry and analytical work \textcolor{blue}{(0.25 columns)} \\
    \subsubsection{Stochastic geometry analytical models}
    % \textcolor{blue}{If fictitious models are simple, they allow  analytical evaluation/model of LOS probability. Advantage: closed-form equations that are tied to building model parameter. Drawback: only works for very simplistic assumptions. Some examples include:} \\
    If fictitious building coverage models are simple, they allow analytical evaluation and modeling of the \gls{LOS} probability. This leads to closed-form equations tied to building and coverage parameters \cite{geometryUAV-alhourani}, but only work for very simplistic assumptions that do not match reality, e.g., cylindrical buildings, equal building size, and assumptions on building location distribution.
        % \begin{itemize}
            % \item Work in \cite{pang2021heightdependent}, uses the same assumptions as ITU-R model \\
            % \textcolor{blue}{}
            % \item Work in \cite{alhourani-stochgeom} derives an analytic expression for LOS probability in urban environments using stochastic geometry \\
    Some examples of such work are \cite{geometryUAV-alhourani} and \cite{alhourani-stochgeom} where simplistic assumptions were made on building density distribution (point Poisson process) and building sizes (cylindrical buildings with the same radius), and provided a closed-form \gls{LOS} probability model. Ref. \cite{Pasi} considers only the impact of clutter such as trees on the \gls{LOS} probability, but not buildings. Another example of such a stochastic geometry model is the work in \cite{los-uav-mmwave-heath}, where the authors studied millimeter wave \gls{UAV} \gls{LOS} probability using Manhattan line processes for the urban grid layout and assuming certain building height distributions and derived closed form \gls{LOS} probability for \gls{UAV} communications.
            % \item Drawbacks: Simplistic assumptions (\cite{alhourani-stochgeom}) such as cylindrical buildings with point Poisson process distribution and no streets or Manhattan grid-like environment. \\            
    % \item Some work uses a mixture of methods
    % \item Machine Learning work: \\
    \subsubsection{Machine learning-based models}
    % \textcolor{blue}{Finally, another category of works uses neural networks to model the LOS probability extracted from geospatial or simulation data. While this can be useful, it is not a stochastic model but more of a deterministic prediction of the LOS in each cell, so it has a different goal than our current work.} \\
    Finally, another category of works uses neural networks to model the LOS probability extracted from geospatial or simulation data. The data source in such cases is either ray tracing or synthetic simulation data as discussed previously. While this can be useful, it is not a stochastic model but more of a deterministic \textcolor{black}{site-specific} prediction of the LOS in each cell. 
    \textcolor{black}{One can train a neural network to predict LOS probability at particular distances in a specific cell, such as in \cite{ml-los2},\cite{a2g-ml}. This is different from deriving a stochastic model that describes the behavior of \gls{LOS} probability in different environment. One could use a machine learning model to generate a set of \gls{LOS} probability curves for a particular environment, but questions arise as to the interpretability and which combination of input parameters to use to generate realistic curves. On the other hand, in a stochastic model, which is the aim of this paper, the model parameters have direct physical interpretability as we will discuss below; the selection of input parameters reduces to sampling their distribution while considering correlations (as we will show below). We stress that our work is not a replacement of machine learning models but is parallel to them and allows for more interpretable insights into the general behavior of LOS probability and its parameters. For applications where the goal is to maximize site-specific prediction accuracy, then machine learning is very well suited for that.}
    However we highlight some notable machine learning results.
        % \begin{itemize}
            % \item Work in \cite{ml-sim} uses a neural netowrk approach based only on urban geospatial data and raytracing, they find large LOS probability is rather large at closer distances (as observed in \cite{Rappaport}) and even at the cell-edge. Moreover, the 3GPP $d_1/d_2$ model is not a good fit even after optimizing the $d_1$ and $d_2$ parameters. However, the LOS was not studied on the street level, as explained above. \\
            The authors in \cite{ml-sim, ml-uav} and \cite{a2g-ml} observed that the LOS probability in practice is larger than what 3GPP predicts (similar to what was observed in \cite{Rappaport}), both at intermediate distances (around 200m-400m) and at the cell-edge. Moreover, the 3GPP $d_1/d_2$ model was not a good fit even after re-fitting the $d_1$ and $d_2$ parameters to their geospatial data.\\
    Based on all the above, we can summarize the drawbacks of the available models for macrocells as follows.
     \begin{itemize}
        % \item 3GPP model and similar are based on small number of measurements/ray tracing, and severely restricted class of environments: "typical" European city where generally there are no buildings greater than 4 storeys, hence a having a BS height of 25m higher than the surroundings. \\
    \item The 3GPP model and similar models are based on a small number of measurements and ray tracing simulations in a restricted class of environments, namely, "typical" European cities where generally there are no buildings greater than 4 stories, and the BS height is around 25m.
    % \item American (and Asian) cities (and even some European cities) have different built-up, but were ignored in model construction. As we will show, this leads to very different LOS probabilities. \\
    American cities (as well as Asian and some European cities), which have different built-up characteristics, building density and heights distributions, were ignored in constructing the 3GPP model. As we will show later, this leads to very different LOS probabilities. 
    % \item None of the previous work used a large scale national-level database to derive their models; thus accuracy is doubtful \\
    \item Furthermore, none of the previous works, including the ones using geospatial data, have used a large- scale nation-level database as the basis for deriving their models; thus their general applicability is doubtful.
    
    % \item Possibly because of small number of measurements, it  is assumed that each cell in the environment has the same LOS probability. However, even within a continent/country, there are significant difference both in the BS height distribution in urban environments and in the building heights in those environments; this again leads to different LOS probabilities than predicted as we will show in our results. \\
    \item Existing works assume that each cell in a particular environment class follows the same LOS probability curve. However, even within the same class in the same country/continent, there are significant variations in both the BS height distributions and building heights. This again leads to different LOS probabilities than predicted, as we will show later in this paper.
    
    % \item no distinction between high-rise urban (that we call metropolitan in this work) and other types of urban environments from a PROPAGATION point of view, \\
    \item Several important environmental categories are not represented at all, or mixed together with other categories. Specifically, there is no distinction in the urban models between areas with high-rise and skyscraper buildings (which we call metropolitan environments in this work)  and other types of urban environments, though these differ significantly from a propagation point-of-view.    
    % \item Models are for urban and rural only, no dedicated model for suburban which is a large portion of the US, \\                
    Furthermore, there is a lack of a model for suburban areas in the US, which is an environment that constitutes a large portion of populated areas, yet from a propagation point of view is distinct from either urban or rural environments.
    % \item some works are purely analytical and based on unrealistic assumptions (no streets, cylindrical buildings with constant radii, distributions of heights that don't necessarily match reality), \\
    % The assumptions used in deriving the models, whether from simulation, geospatial databases or using purely analytical methods, are lacking as explained above.
    % \item some work does not incorporate STREET level LOS probability even in urban environments 
    % \item None of the works model the parameters of their model statistically to be general enough for the type of environment under study. \\
    % Finally, no work models the distribution of the parameters of the LOS models, again due to the small number of data points.
     \end{itemize}
    
    % \item Current work closes this gap: \textcolor{blue}{(0.25 columns)} \\
    \subsection{Contributions}
        Our current work closes these gaps. We provide both a methodology on how to categorize and evaluate \gls{GIS} databases \textcolor{black}{that can be used on any other region in the world}, we propose better models (more general fitting model, cell-by-cell modeling), and we provide fully parameterized models for the whole US. Specifically, our contributions can be summarized as:
        \begin{itemize}
            \item We provide a complete framework for classifying radio environments into different categories: rural, suburban, urban and metropolitan. This approach takes both building height and building density into account.  
            \item We use an efficient method for evaluating LOS existence from geospatial data, and include details about environmental and geospatial data reliability considerations. We use real-world environment data bases and BS deployment data spanning the entire continental US, evaluating  more than 13,000 BS locations. To our knowledge, it is the first time that real-world datasets of this size have been used. 
            \item  We show that LOS probabilities should be modeled cell-by-cell, not just averaged over all cells, and demonstrate that for outage and \gls{SIR} evaluations. This is a fundamentally different way of modeling LOS probability that reflects the diversity of city buildups.\footnote{ Our conference version \cite{globecom-los} introduced this approach, though evaluated only with Los Angeles-based data.}
            \item We model LOS probability in each cell as a parametric model, and the parameters as correlated random variables. We provide (parametric) \glspl{CDF} of these parameters and their correlations, for each environment. We provide full parametrization of this model for US-based rural, suburban, urban, and metropolitan areas.
            \item For backwards compatibility and comparability, we also propose a variation of the 3GPP $d_1/d_2$ model based on the average cell behavior.
            % \item We compare the performance of our models relative to 3GPP models in terms of outage probability, and \gls{SIR} distributions.
            
        \end{itemize}       
        % \item Organization of the rest of the paper \textcolor{blue}{(3-4 lines at most)} \\
        The rest of our paper is organized as follows. Section \ref{sec:LOS} discusses the employed datasets and the algorithms used for extraction of the LOS. We present the environment classification of BSs and special considerations in Section \ref{sec:Env-class}. The proposed model is outlined in Section \ref{sec:Model} and the processing of the LOS data and fitting procedure are explained in Section \ref{sec:Fitting}. Then we show the results of our fitting, the statistical modeling of the model parameters per environment, and outage simulations results in Section \ref{sec:Results}. Finally, we conclude and present possible future extensions to our work in Section \ref{sec:Conclusion}.
        
        % \begin{figure}
        %     \centering
        %     \includegraphics[width=0.4\textwidth]{Figures/Tower 18 Single Starting Point.png}
        %     \caption{Caption}
        %     \label{fig:my_label}
        % \end{figure}            
    
\section{LOS Definition and Extraction Methods} \label{sec:LOS}
% \textcolor{red}{(around 2 columns)}
% \begin{itemize}
  \subsection{LOS Definition}              
    % \begin{itemize}
        % \item Optical LOS and distinction from frequency dependent definition \textcolor{blue}{($\leq$0.5 columns)} \\
        In this paper, we define the LOS as the optical connection between BS and street point \textit{on the ground}, meaning if there exists an uninterrupted direct line between BS and street point on ground level, the point is considered a LOS point. While other definitions based on the Fresnel zone \cite{Pasi} exist, those are frequency dependent and would thus require either models for specific bands, or a more complicated modeling approach. Furthermore, we do not consider cluttering objects such as trees and vegetation, which are not contained in digital databases at a national scale. If their statistics are known for a particular area, the techniques of \cite{Pasi} can be used to assess their impact. Note that other definitions of \gls{LOS} based visibility regions exist and are used in site-specific cases \cite{cost259}. \textcolor{black}{The choice of ground level height for the street points is because we analyze \gls{UE}s with heights less than 1.5m (since most smartphones are not at ear height anymore, but lower). We compare the effect of a greater street point height, namely 1.5m, on the \gls{LOS} probability in Sec. VI; as we show there, the effect of the increased height is not very significant.} 
        %\item Area and geographic diversity, never been done before

        % \item Algorithm for extracting LOS (GIS) \textcolor{blue}{(1 column)}: Data source, software used, algorithm inputs, outputs and general procedure
    % \end{itemize}    

   \subsection{Data sets and Extraction algorithm} 
    % \begin{itemize}
        % \item Data set of BS. Stress it is only MACROCELLS.
        The dataset we use containing BS information: coordinates, height above ground, and height above sea level, contains macrocells only. It is extracted from the \gls{FCC} database \cite{fcc}, and contains a total of 160,623 BSs. The information was augmented with a geospatial building database combining terrain elevation and building information.
        
\begin{figure}[t]
          \centering
          \subfloat[OSM data\label{fig:bs-ex-osm}]{
            \includegraphics[width=0.4\textwidth]{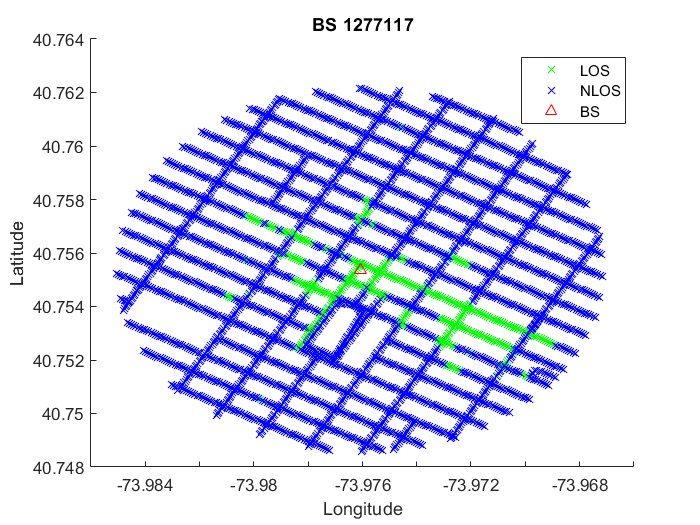}
          }\hfill
          \subfloat[MS data\label{fig:bs-ex-ms}]{
            \includegraphics[width=0.4\textwidth]{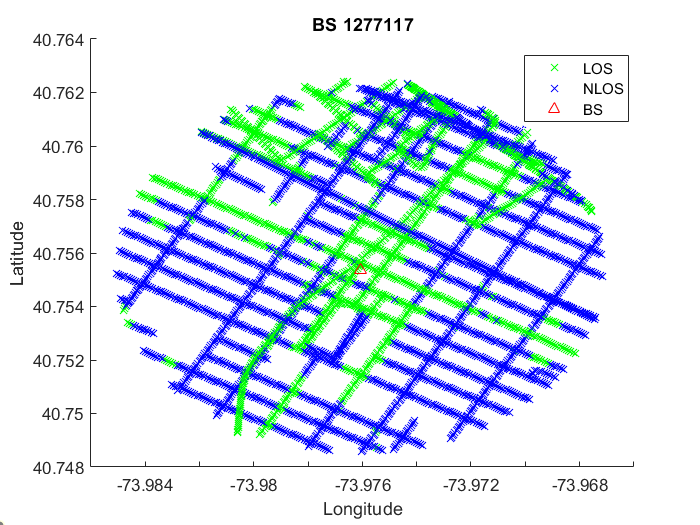}
          }
          \caption{2D LOS scatter plot comparison between (a) OSM dataset and (b) MS dataset of BS~1277117.}
          \label{fig:dataset-comp}
        \end{figure}

        % \item Different geospatial datasets available. Need to use public-domain. Two most prevalent are MS, OSM: MS has less footprints but more heights, we found that footprints are severely lacking in metropolitan areas (NY, LA), not usable \textcolor{blue}{(1-2 columns if figures added)} \\
        For the terrain elevation, we use a \gls{DEM}, in particular the 1/3rd arc-second \gls{DEM} updated by the \gls{USGS} \cite{usgs}.
        For the height and location of the buildings, there are multiple possible sources for the dataset. We restrict our attention here to public-domain databases, the two most prominent of which are \gls{OSM} \cite{geofabrik} and \gls{MS} US building footprints dataset \cite{ms-ds}. The OSM dataset is an open-source database of geospatial information which includes building information, roads, rivers, and more. The OSM database is crowd-sourced, i.e., the public can upload information. For this reason, the level of detail and completeness of information can vary greatly between some areas; generally remote and low population areas have fewer data available, while urban and well-populated areas tend to have more complete information. The \gls{MS} dataset is generated by a deep learning model trained to detect two-dimensional building footprints from high-resolution satellite imagery. The dataset does not typically contain building heights, and whatever height information available is very sparse. Moreover, according to our analysis, the \gls{OSM} dataset is a better representation of the real environments, as it contains significantly more footprints in key urban areas, including but not limited to downtown Los Angeles and Manhattan. A sample comparison of the data is shown in \figurename \ref{fig:dataset-comp} for a BS in Manhattan. We see that use of the MS data results in considerably higher number of LOS locations, which (as further detail investigation showed) is an artifact related to the missing footprints and height information of the data. Consequently, we used the \gls{OSM} dataset in our investigation. In particular, the dataset is generated based on a mixed raster-vector method that incorporates the DEM as discussed above.  The building features, including height if available, are extracted from OSM vector layers from a Geofabrik server \cite{geofabrik}. The street features are obtained from TIGER/Line geo-databases \cite{tiger}. %For the evaluation of the MS dataset, we ....}
        
        % Figure
        
        % \item OSM dataset shortcomings: Open source data, not all of the buildings in the database, more footprints less heights. Pros: most environments are accurately represented when comparing to Google Earth, height-footprint ratio is still high (filtering will be done) \textcolor{blue}{(0.5 columns)} \\
        Although the OSM dataset is better, it is still not complete in terms of heights for all the available footprints, which is a drawback due to its crowd-sourced open nature. This causes inaccuracies while computing the LOS status of street points, causing an artificially high LOS probability that is not reflected in reality. For that reason, further statistics on building coverage and height-to-footprint information in each cell must be studied to filter out cells with incomplete data. This will be addressed Sec. \ref{sec:Env-class}.

        % \item Algorithm for extracting LOS (GIS) \textcolor{blue}{(1 column)}: software used, algorithm inputs, outputs and general procedure \\
        The street points are sampled in the middle of every road in 5m intervals in a 1km radius around the BS. Ray tracing  is used to determine the LOS visibility of the street point: we draw a \gls{3D} line between the BS and the street point, sample it, and compare the height of the points on the line to the heights of the buildings intersecting this line. If the former is greater than the latter everywhere, the point is \gls{LOS}, otherwise it is \gls{NLOS}.

\begin{figure}
            \centering
              \subfloat[BS~1063140 scatter plot\label{fig:la-tower-cesium}]{
                \includegraphics[width=0.4\textwidth]{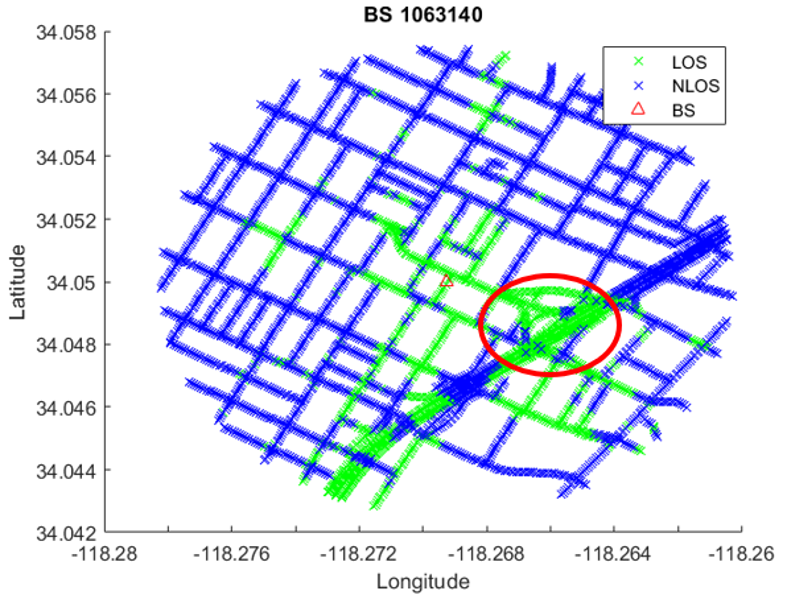}
              }\hfill
              \subfloat[Ray-tracing check in Cesium\label{fig:cesium-plot-la}]{
                \includegraphics[width=0.4\textwidth]{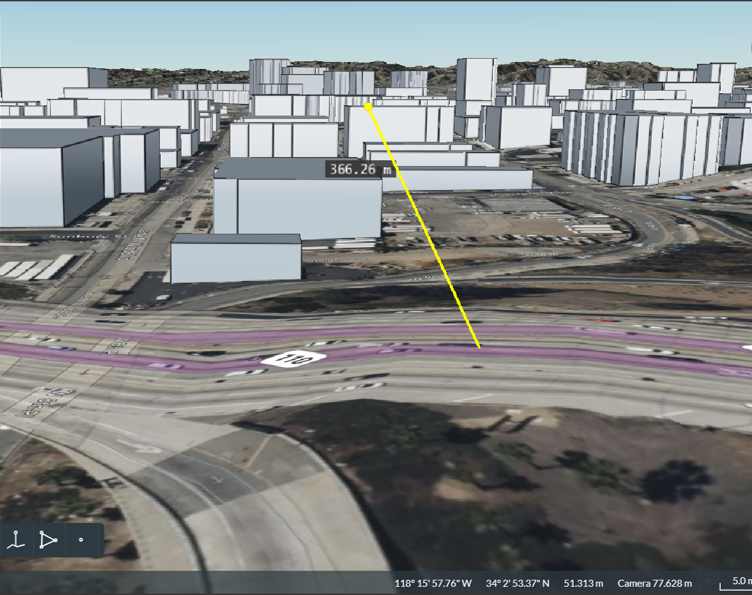}
              }
            \caption{Example of verification of the LOS for (a) BS 1063140 as in the OSM dataset and (b) the red circled area's physical location using ray tracing in the Cesium tool}
            \label{fig:cesium-comp}
        \end{figure}

        % \item State that we manually verified some towers through an online database and manual ray drawing (Important justification) \\
        To ensure the correctness of the algorithm as well as the database, we manually verified street point visibility for some \gls{BS}s in metropolitan areas of downtown Los Angeles and Manhattan by using the Cesium tool \cite{cesium}. This allows us to interact with Google Earth data overlayed with OSM building models and draw lines between any \gls{3D} point, in this case between \gls{BS} and street point to check if it intersects any building, hence verifying the correctness of the LOS label for the point. We perform this check for areas that are distant from the \gls{BS} where one would usually not expect to find a high number of \gls{LOS} points. The verification using Cesium was in line with the LOS labels extracted from the OSM database.
        One verification example for a downtown Los Angeles area is shown in \figurename \ref{fig:cesium-comp}
        
        % \item Limitations of our data-based approach \textcolor{blue}{(0.5-1 columns)}: Publicly available geospatial datasets. Towers with low building coverage percentage and low building-height-to-footprint ratio needed to be filtered. \\
        To summarize, publicly available building-height databases that allow automated evaluation show limitations in accuracy. While the OSM dataset is the best of the available ones, further filtering is required to increase the reliability of the results; this will be discussed in more detail in Sec. \ref{sec:Env-class}
    % \end{itemize} 

        % \item Height data set: Source and extraction method \textcolor{blue}{(1 columns)}            

    % \end{itemize}    

\section {Environment classification} \label{sec:Env-class}
    % \textcolor{red}{(4.5 columns)}
    % \begin{itemize}           
                 
    % \textcolor{blue}{Considered environmental classes: rural macrocell (RMa), suburban macrocell (SMa), urban macrocell (UMa), and metropolitan macrocell (MetMa). Different criteria for distinction available: inter-site distance,antenna heights,  average building heights in the cell, building density.}
    We consider 4 categories of environments in our work: \gls{RMa}, \gls{SMa}, \gls{UMa}, and \gls{MetMa}. There exists different criteria for classifying the environments, those include  antenna heights, building density, average building heights in the cell and \gls{ISD}. We discuss these criteria below as well as our own proposed criteria.

    % \item Why ITU-R classification based on antenna heights and inter-site distance is not a good criterion: operator dependent, buildings affect propagation conditions more than antenna heights. \textcolor{blue}{(1 column)} \\
    \subsection{Possible classification criteria}
    The ITU-R recommends using \gls{ISD} and antenna heights to classify the BSs and assumes an increasing antenna height for each category. These criteria, however, are an operator-dependent choice in practice. \gls{ISD} depends on the amount of service usage in a particular area, which has nothing to do with characterizing the propagation, and can change with time. Antenna height might depend on where space can be rented; often we find antennas to be mounted on building rooftops that have a varying height even in the same category of environments. Moreover, the radio propagation and \gls{LOS} on the street level also depends on the height of the obstructing buildings. While building density is usually a differentiating factor between types of environments, it is not as impactful as building heights when we are considering street-level \gls{LOS}, since the optical connection between \gls{BS} and street-level \gls{UE} is considered. We presented a classification based on average building heights in a cell in our conference version \cite{globecom-los}.
    %which is not taken into account in any previous work.

    %%%%%%%%%%%%% NEW: %%%%%%%%%%%%%%%%%%%%
    Additionally, although the ITU prescribes parameters such as building heights in \cite{itu-r-rec} for a typical environment type, they do not provide a classification method that allows to distinguish environments. Furthermore, the ITU parameters do not take into account various important factors such as building footprint geometry. As a matter of fact, Section 2.1.5 of Ref. \cite{itu-r-rec} acknowledges that since the building density and heights vary greatly from one region to the other, its proposed statistics do not fit every environment.
    %%%%%%%%%%%%% END NEW: %%%%%%%%%%%%%%%%
    
\begin{figure}
            \centering
                \subfloat[Large-height industrial urban cell\label{fig:misclass-ex-urban}]{
                    \includegraphics[width=0.4\textwidth]{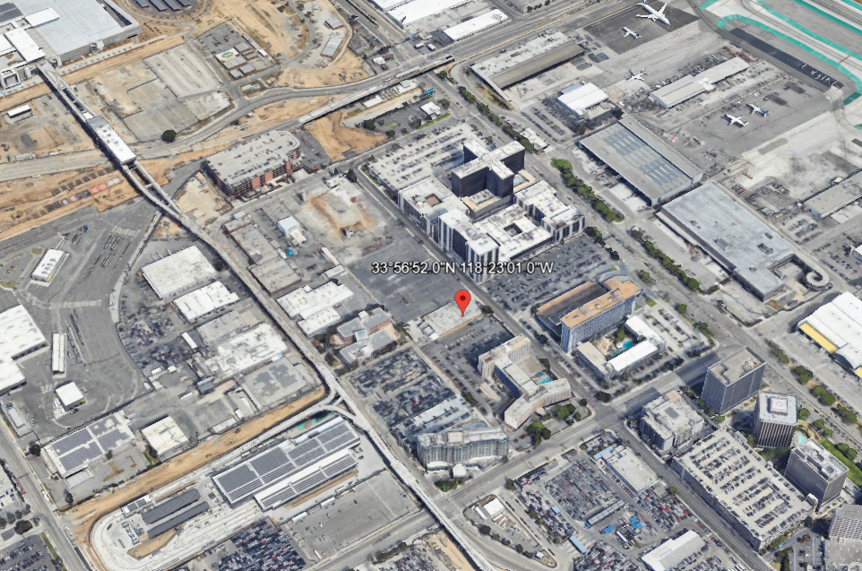}
                  }\hfill
                  \subfloat[Lower-height industrial urban cell\label{fig:misclass-ex-urban-suburban}]{
                    \includegraphics[width=0.4\textwidth]{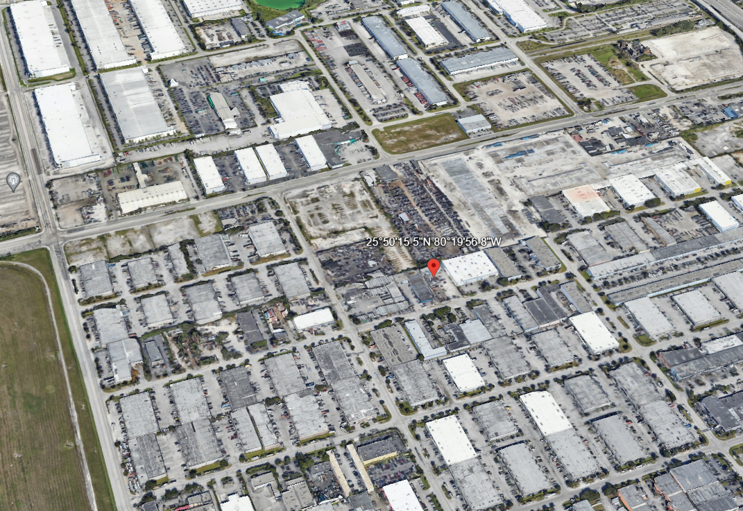}
                }
            \caption{Comparison of two urban cells, where (a) is classified as urban and (b) is classified as suburban from a propagation point-of-view}
            \label{fig:misclass-ex}
    \end{figure}

    % \item Discussion of some caveats to using average building heights: Example of industrial urban with low density and large average building heights vs. industrial urban with high density and low building heights (misclassification as urban) \\
    The density criterion comes with a limitation: some cases of industrial urban areas with high building density and low building heights can be classified as suburban from a \textit{radio propagation} point-of-view. This might seem counter-intuitive when looking at the usage of the buildings. However, from a propagation point-of-view, there is only a small difference between the heights and obstructions encountered in this type of area and those encountered in a densely packed suburban area.
    % \item Figures of a "suburban" industrial area vs true urban industrial \\
    \figurename \ref{fig:misclass-ex} illustrates the difference between two types of industrial urban environments

\begin{figure}
        \centering
        \includegraphics[width=0.4\textwidth]{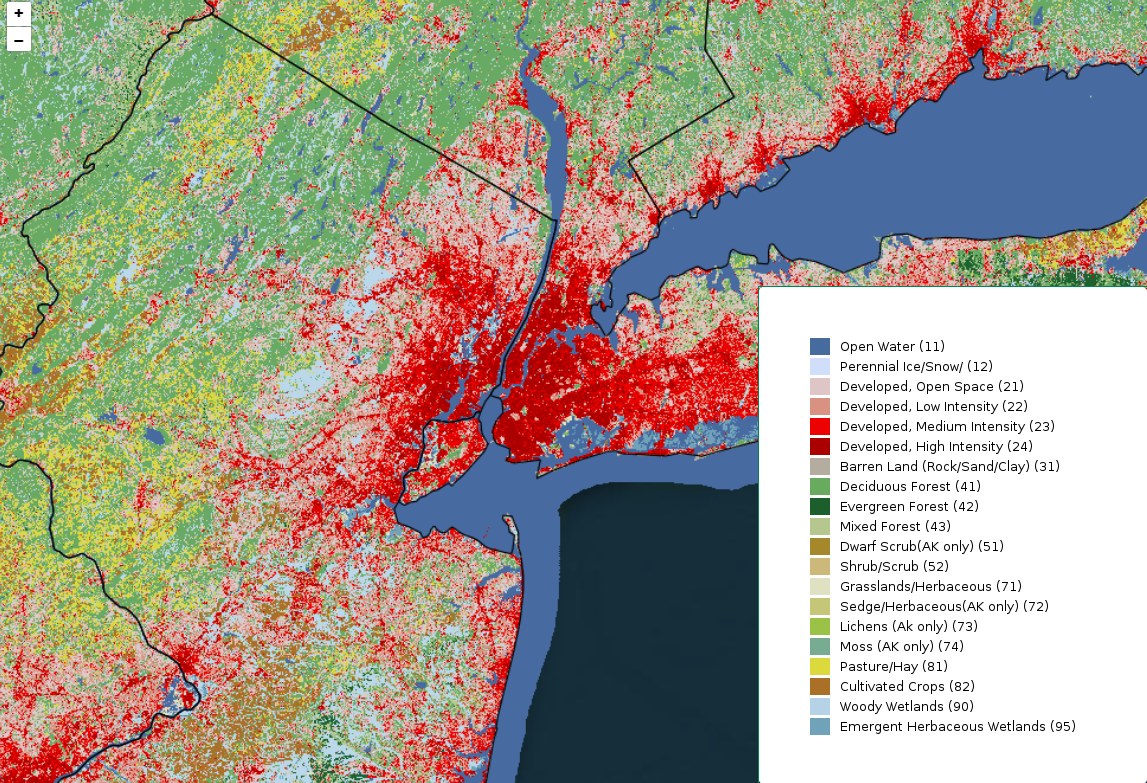}
        \caption{\gls{NLCD} map for the New York area}
        \label{fig:nlcd}
    \end{figure}

    % \item NLCD classification: density based, does not take into account building height \textcolor{blue}{(0.5 column)} \\
    Another classification criterion that could be used is the land coverage percentage, as in the \gls{NLCD} \cite{mrlc}. This dataset provides information on type of land cover in the US, in 8 categories (and their subcategories) ranging from water, forests, to developed urban areas, as in Fig. \ref{fig:nlcd}. One could use the subcategories and combine them to get our 4 categories, however this assignment would be more arbitrary than using average building heights since it is not based on numerical threshold but is categorical instead. Moreover, the data is not updated on a yearly basis but periodically on a longer time frame, and since the dataset is based on land coverage it suffers from the same drawback as using building coverage (density) criteria.

    % \item Our criteria are based on \cite{globecom-los}: average building heights, and coverage only used for rural areas distinction due to the similarity of average heights to those in lower coverage suburban areas \\

\subsection{Proposed classification criteria}
Consequently, and similar to the conference version of our work \cite{globecom-los}, we propose classification criteria based on average building height in the 1km-radius cell. The thresholds for each category are based on \cite{heights-class} where the authors used the number of floors to define their categories. We also found that adding a criterion on coverage improves the classification of rural cases. It is important to note that there is no objective ''ground truth" data against which the classification results can be compared, because the thresholds for discrimination are inherently arbitrary, and behavior does not change abruptly when crossing a threshold.\footnote{To quote George Box: ''All models are wrong, but some are useful."} \textcolor{black}{As a matter of fact, there is no ''official" definition of what a specific type of environment, e.g., urban, is - even manual inspection of satellite pictures would still rely on the subjective assessment of the viewing person (not to mention that this would be  impractical to do for the entire dataset of over 160,000 cells).} \textcolor{black}{To the best of our knowledge, 3GPP does not provide explicit classification criteria for environments; specifically, it does not provide quantitative criteria of how a particular environment on a map should be classified. In terms of the model, it postulates a particular \gls{BS} height and inter-\gls{BS} distance, but does not consider other environmental parameters. In terms of the measurements on which the 3GPP model parametrization is based, they are based on some very clearcut and special cases, but again do not provide any quantitative criteria of where the boundary is between different environment classes; furthermore the measurements are done only in a small number of locations that do not consider the diversity of environments within one class.} We do a trial-and-error process for verification \textcolor{black}{of our classification criteria} while comparing key urban, metropolitan, suburban and rural areas (e.g., Manhattan and surrounding areas, metropolitan Los Angeles area) with Google Earth geography. Table \ref{table:env-stories} shows the criteria for each environment.
    % \begin{figure}
    %     \centering
    %     \includegraphics[width=0.4\textwidth]{Figures/MRLC-NLCD.png}
    %     \caption{Land coverage data map for an area of Southern California}
    %     \label{fig:nlcd}
    % \end{figure}            

    % \item Table showing the criteria for each environment \textcolor{blue}{(0.5 column)}
    
    \begin{table}[]
        \centering
        \caption{Environment classifications and average building heights}
        \label{table:env-stories}
        \begin{tabularx}{0.5\textwidth}{|>{\centering\arraybackslash}X                
        |>{\centering\arraybackslash}X|}
             \hline
             \textbf{Environment} &  \textbf{Criteria}\\
             \hline 
             RMa & Average Building Heights: 0-2m \& Building Coverage $< 10\%$\\
             \hline 
             SMa & Average Building Heights: 2-10m\\
             \hline
             UMa & Average Building Heights: 10-25m\\
             \hline
             MetMa & Average Building Heights: $\geq$ 25m\\
             \hline
        \end{tabularx}        
    \end{table}

\begin{figure}
        \centering
        \includegraphics[width=0.4\textwidth]{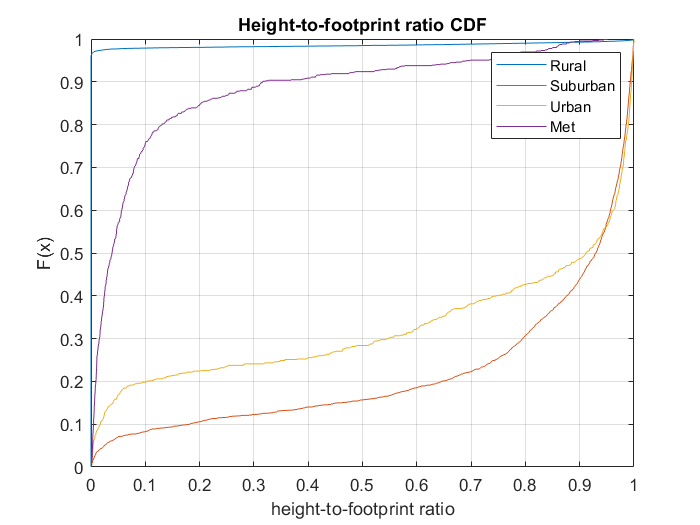}
        \caption{Height-to-footprint ratio distribution for all environments before filtering}
        \label{fig:hfi-cdf}
    \end{figure}

    Due to the crowd-sourced nature of the data, not all building information is present for all areas. In some cases, information about buildings might be missing completely, while in others, the information about the footprint might exist but the height information is missing. Thus, to increase reliability, we use the distribution of building coverage area in each cell and the height-to-footprint ratio statistics as further criteria to filter and classify the cells. The building coverage area is defined as the ratio of total area occupied by buildings in a cell (1km radius around the BS, see below) to the total area of the cell, and the height-to-footprint ratio is defined as the ratio of the number of building footprints that have height information to the total number of footprints. We use the building coverage statistics for the environmental classification of the BSs only, since a particular class of environment can have different building densities, as shown in \cite{Townend-conf}. This is done by classifying all towers with building coverage less than 10\% as rural, which has been observed to improve our classification. The height-to-footprint ratio is used as the filtering criteria to keep only the most reliable BSs; it is a criterion that has not been studied in any previous work. We use the threshold of 90\% for the height-to-footprint ratio to keep only the BSs with height information greater than 90\% in all environments. The distribution of the height-to-footprint ratio for all environments before filtering is shown in Fig. \ref{fig:hfi-cdf}.
    % the cells are subsequently filtered based on their building coverage distributions and height-to-footprint ratio distributions. Those are defined respectively as the ratio of total area occupied by buildings in a cell (1km radius around the BS) to the total area of the cell, and the ratio of the number of building footprints that have height height information to the total number of footprints. 
    % We use the 90-th percentile as a threshold for filtering and keep all the cells in each environment that have coverage and height-to-footprint ratios greater than the 90-th percentile.\\
    %
    Note that one could try interpolating the missing building heights using an average of the neighboring buildings' heights, this approach only resulted in an average of 1.4\% change in the LOS labels of the measurement points for the 5237 cells where we tried this interpolation.
    
         % Not sure if the filtering by coverage is necessary, since we obtained plausible results wihtout it and Townend didn't do any filtering.
    % \begin{table}[]
    %     \centering
    %     \begin{tabularx}{0.5\textwidth}{|>{\centering\arraybackslash}X
    %     |>{\centering\arraybackslash}X
    %     |>{\centering\arraybackslash}X|}
    %          \hline
    %          \textbf{Environment} & \textbf{Height Classes} & \textbf{Average Building Height}\\
    %          \hline 
    %          RMa & 1 or less stories & 0m to 3m\\
    %          \hline 
    %          SMa & Low, Low-medium & 3m to 12m\\
    %          \hline
    %          UMa & Medium, medium-high, high & 12m to 36m\\
    %          \hline
    %          MetMa & Very high & 36m or more\\
    %          \hline
    %     \end{tabularx}
    %     \caption{Average building height classification into environmental classes}
    %     \label{table:env-stories}
    % \end{table}
                
    % Already addressed above
    %\item This is only MACROCELL 
    
    % Already done in previous point \item Figure: Include example of urban and suburban: Different geographical locations from Google Earth \textcolor{blue}{(0.5 columns)}
    
    % \item State that we analyze the building height distribution based on the above classification and the correlation of building heights and antenna heights with U and F parameters (the results and figures for this analysis will be in the results section).
    After the classification, we study the statistical distribution of the LOS parameters in each environment as well as their correlations.
    
% \end{itemize}

\section{LOS Model} \label{sec:Model}
    % \textcolor{blue}{(1 column)} \\
    % \textcolor{blue}{This section explains our proposed LOS model. The key differences between our model and 3GPP are the cell-by-cell approach, but we also introduce a new. Within each cell, we use a modification of 3GPP model. }
    % \begin{itemize}
    % \item Existing 3GPP models: different models of LOS probability for different scenarios \cite{38901}, with the urban macrocell (UMa) and $d_1/d_2$ models being the most focused on. 
    
    % \item Models for urban, rural, etc. check the table from 38.901. 3GPP also defines models for rural and suburban areas (0.5 columns table + 1 column discussion)

    % \item ... \\
    While 3GPP has defined several models, the UMa model receives the most focus. It is based on the $d_1/d_2$ model and later extended to include endpoint height \cite{38901}. If the endpoints (or UEs) are at street level as in our case, it reduces to:
     \begin{align}
    p_{\rm LOS,3GPP}(d) = \mathrm{min}(\frac{d_1}{d},1)(1-e^{-d/d_2}) + e^{-d/d_2} ~~.
    \end{align}

    % \item Observed discontinuity in LOS between constant probability region and decaying region, as well as larger LOS at closer distances similar to what some literature observed (as mentioned above) that was not modeled \\
    \subsection{Model formulation}
    We observed in our data a discontinuity between the constant probability region and the exponentially decaying region that cannot be eliminated by reducing the sampling distance. This discontinuity is not always severe, but depends on the distribution of the LOS points in the cell area: the more uniformly spread they are, the less discontinuity we observe. A possible physical explanation is a cutoff of optical LOS by a dominant building (or buildings). 
    
    % \item Figure of scatter plot vs PLOS plot \\
    \begin{figure}[t]
      \centering
      \subfloat[BS 11 LOS points\label{fig:tower11scatter}]{
        \includegraphics[width=0.4\linewidth]{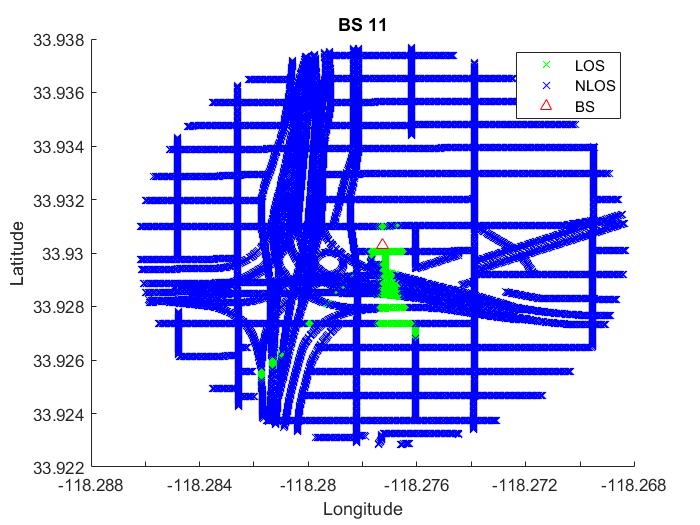}
      }\hspace{0.01\textwidth}
      \subfloat[BS 11 LOS probability\label{fig:tower11plos}]{
        \includegraphics[width=0.4\linewidth]{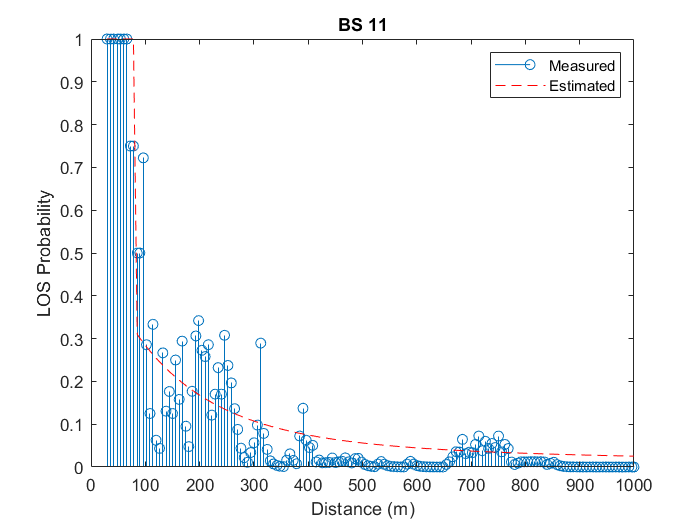}
      } \\[0.4em]
      \subfloat[BS 23 LOS points\label{fig:tower23scatter}]{
        \includegraphics[width=0.4\linewidth]{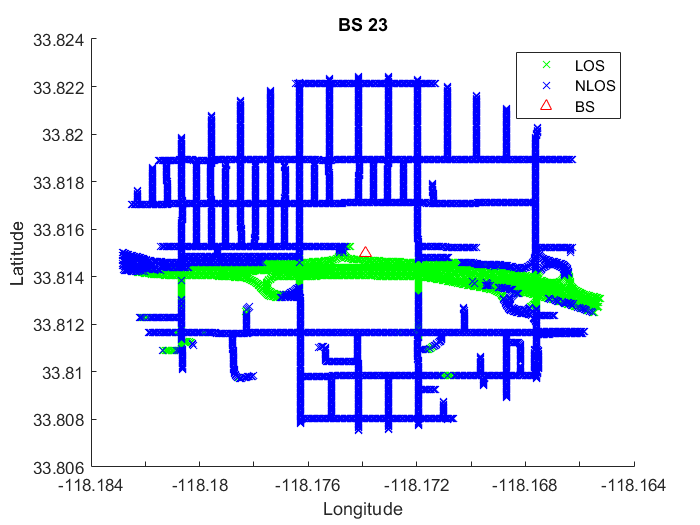}
      }\hspace{0.01\textwidth}
      \subfloat[BS 23 LOS probability\label{fig:tower23plos}]{
        \includegraphics[width=0.4\linewidth]{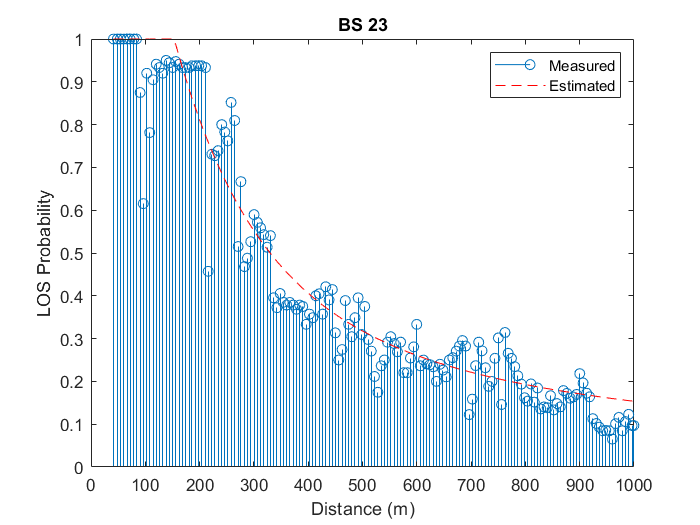}
      }
      \caption{Illustration of discontinuity in LOS probability. Scatter plots of spatial LOS for BS 11 (a) and 23 (c) and the corresponding empirical LOS curves (b) and (d).}
      \label{fig:scatter}
    \end{figure}
    
    % \item Our new model: F parameter to model discontinuity due to uneven concentration of LOS points around BS. \\
    The effect is illustrated, for a specific cell, in \figurename~\ref{fig:scatter} along with the resulting LOS probability and fit, as explained next. The discontinuity motivates us to  introduce a new parameter to the 3GPP model that we call the ``scale" or discontinuity parameter $F \in [0,1]$ that multiplies the exponentially decaying part.
    % \item Equation for our LOS model \\
    Thus, the equation for our model is:
    \begin{align}
    p_{\mathrm{LOS}}(r; U,W,F) = 
    \begin{cases}
        1 & r \leq U\\
        F(\frac{U}{r}(1-e^{-\frac{r}{W}}) + e^{-\frac{r}{W}}) & r>U
    \end{cases} \label{equ:plosmodel}
    \end{align}
    where $F$ is the scale parameter, $U$ is the cutoff distance, and $W$ is the decay parameter.
                      
    % \item Average model explanation: to describe the LOS probability in a particular environment on average, and to have similar specifications as the 3GPP where one model is given per environment, allows for a fair comparison.(note that fitting to the average probabilities is different from averaging over the cell-by-cell parameter distributions) \\
    
    \subsection{Average model}
    The formulation in (\ref{equ:plosmodel}) is used to fit the cell-by-cell (also referred to as the ``ensemble" model since it can be used to generate an ensemble of cell-by-cell curves) as well as the average environmental LOS probabilities. Average models are obtained by combining the data of all cells belonging to the particular environment, creating a hypothetical ``typical" cell, then fitting the data to the model as we will outline in the next section. Note that this average model is different from taking the average of the $U, \: W, \: F$ distributions for each environment, since it represents what a typical or average environment would look like from a geographical buildings layout perspective.        
    % \item Does NOT always get "smeared" for average model. \\
Note further that while the physical motivation for the scale parameter is mainly tied to a {\em per cell} investigation, we observed that the discontinuity is still present in the {\em average} models for each environment; however it is less pronounced in that case, due to the increased homogeneity of the environment in the typical cell created by combining the data from multiple towers.

    % \end{itemize}
    
\section{Data processing and model fitting} \label{sec:Fitting}
% (2.5 columns)
% \begin{itemize}
    % \item Binning algorithm explanation with equations, transforming 2D data to 1D. \textcolor{blue}{(0.75 columns)} \\
    We analyze the LOS probability within a distance of 1 km of the BS. While in most cases (except rural areas) this may be larger than the cell radius, the information for the larger distances is useful for interference assessment. 

    \subsection{Binning}
    To transform the two-dimensional (2D) data into distance dependent data, we proceed by binning points within a certain range together and computing their average distance and empirical LOS probability. \textcolor{black}{Note that this binning will naturally average out the  location-specific impact of the buildings, but this is a general feature of probabilistic modeling of \gls{LOS}}. Since the spatial resolution of the analyzed data is 5m, we choose a distance bin width of 5m as well, i.e., for each bin 
    \begin{align}
        r \in [r_{l,i}, \: r_{l,i} + 5], \quad  i=1, \dots 200
    \end{align}
    where $r$ is the 2D distance from the street point to the BS and $r_{l,i}$ is the lower distance bound of bin $i$, and the relationship between the distance bounds of bin $i$ and $i+1$ is $r_{l,i+1} = r_{l,i}+5$.    
    % \item Equations for computing average distance and empirical LOS probability. \\
    Each bin is represented by the average distance in the bin and the empirical LOS probability, which are computed respectively using:
    \begin{align}
        r_{\mathrm{bin},i} &= \frac{1}{N_{{\mathrm{bin}}_i}} \sum_{j=1}^{N_{{\mathrm{bin}}_i}}{r_j}\\
        p_{\mathrm{meas}}(r_{\mathrm{bin},i}) &= \frac{1}{N_{{\mathrm{bin}}_i}} \sum_{j=1}^{N_{{\mathrm{bin}}_i}}{\mathbb{I_{LOS_j}}}
    \end{align} 
    where $\mathbb{I_{LOS_j}}$ is the LOS indicator of the measurement point $j$, $N_{{\mathrm{bin}}_i}$ is the number of points in bin $i$. Note that the average distance need not be $r_{\mathrm{bin},i}$, but depends on the distance distribution of the points within the bin. 
    
    % \item Log scale fitting of the probabilities to make it easier to fit the $F$ parameter \\
    \subsection{Fitting: performance metrics and algorithm}
    For the fitting of the empirical LOS probability, the goal is to obtain ``reasonable" fits for all distances. We can choose to fit the probabilities on a linear scale by minimizing the \gls{MSE} or fit their logarithms by minimizing the \gls{MSLE}, the latter provides a better fit at larger distances. A \gls{WMSE} approach can be used to achieve the same goal as well; one possible weighting scheme is using the distance of each bin as weight. Another possible scheme is using weights proportional to the inverse of the empirical probability which enforces a good fit at large distances where we expect to have smaller probabilities. 
    The weights for the \gls{WMSE} approaches are given respectively by:
    \begin{align}
        w(i) = r_{\mathrm{bin},i}
    \end{align}
    \begin{align}
         w(i) = \frac{1}{p_{\mathrm{meas}}(r_\mathrm{bin},i)+\epsilon}
    \end{align}
    where $w(i)$ is the weight of bin $i$, and $\epsilon$ is a small value used to avoid infinite weights (we set $\epsilon=0.05$ empirically).
    The formulations of the \gls{MSE}, \gls{MSLE}, and \gls{WMSE} are given respectively by:
    {\small
    \begin{align}
        &\mathrm{MSE} = \frac{1}{N}\sum_{i=1}^{N}{(p_{\mathrm{meas}}(r_{\mathrm{bin},i}) - p_{\mathrm{LOS}}(r_{\mathrm{bin},i}; U,W,F))^2} \\
        &\mathrm{MSLE} = \frac{1}{N}\sum_{i=1}^{N}{( \ln( p_{\mathrm{meas}}(r_{\mathrm{bin},i})) - \ln(p_{\mathrm{LOS}}(r_{\mathrm{bin},i}; U,W,F)) )^2} \\
        &\mathrm{WMSE} = \frac{1}{N}\sum_{i=1}^{N}{w(i)(p_{\mathrm{meas}}(r_{\mathrm{bin},i}) - p_{\mathrm{LOS}}(r_{\mathrm{bin},i}; U,W,F))^2}
    \end{align}
    }
    where $N$ is the number of bins for the cell. The fitting problem is formulated as the minimization of one of the above metrics subject to the following constraints:
    \begin{align}
        & 0\leq U \leq 1000\\
        & W \geq 0  \\
        & 0\leq F \leq 1
    \end{align}
    % \begin{figure}
    %     \centering
    %     \includegraphics[width=\linewidth]{}
    %     \caption{Caption}
    %     \label{fig:enter-label}
    % \end{figure}
    
    % \item Optimization problem formulation with constraints explained \textcolor{blue}{(0.5 columns)}\\ 
    % \textcolor{blue}{We formulate the fitting problem as the minimization of the sum of square errors (SSE) of the linear-scale probabilities on a log-distance scale given certain constraints on the parameters:
    % \begin{align} \begin{split}                
    %     U^*, W^*, F^* &= \argmin_{U,W,F} \sum_{i=1}^{N} (p_{\mathrm{meas}}(r_i^{(\mathrm{log})}) \\ &\quad \quad- p_{\mathrm{LOS}}(r_i^{(\mathrm{log})}; U,W,F))^2\\
    %     &\quad \textrm{s.t.} \quad  0\leq U \leq 1000\\
    %     & \quad \quad \quad W \geq 0  \\
    %     & \quad \quad \quad 0\leq F \leq 1
    % \end{split} \end{align}
    % where $ U^*, W^*, F^*$ are the optimal parameters, $N$ is the number of bins for the cell, $r_i^{(\mathrm{log})}$ is the log-scale distance, $p_{\mathrm{meas}}(.)$ and $p_{\mathrm{LOS}}(.; .)$ are as defined above.
    % }
    We stress that there is no objectively ''best" criterion in general, but that rather the optimum fitting method will depend on the application at hand. 
    
    % \item Grid search and 10 starting points for gradient descent: non convex problem, picking starting point that leads to lowest SSE \textcolor{blue}{(0.5-0.75 column)} \\
    
    Regardless of the chosen metric, the optimization problem is non-convex and non-linear, so to avoid local minima we first perform a grid search to find initial values of the parameters, then use them as the starting points for a gradient descent minimization. Since with a non-convex goal function gradient descent can converge to a local minimum, we use the grid points that give the 10 lowest \gls{MSE} values as different starting points to the gradient descent, and finally select the parameters that give the lowest \gls{MSE} out of the final 10 candidates. We found that using more than 10 starting points does not significantly improve the fitting but increases the running time.

    To choose the best error metric for the fitting, we compare the resulting fits based on average \gls{MSE} over the entire distance range and the average \gls{MSE} over the distance range $r>500m$ for all the \gls{UMa} BSs. The comparison is shown in Table \ref{table:fit-comp} where the \gls{MSLE} shows a better fit for large distances but a slightly worse overall \gls{MSE}, which is an acceptable tradeoff, hence we choose the \gls{MSLE} as the fitting metric. We stress again that there is no objectively ''best" criterion in general, but that rather the optimum fitting method will depend on the application at hand; yet since we aim for a generally applicable model, we select a criterion that gives, in our view, ''reasonable" results. 

    \begin{table}[]
        \centering
        \caption{Comparison of the different possible fitting metrics averaged over the \gls{UMa} BSs.}
        \label{table:fit-comp}
        \begin{tabularx}{0.5\textwidth}{|>{\centering\arraybackslash}X
        |>{\centering\arraybackslash}X
        |>{\centering\arraybackslash}X|}
             \hline
             Metric & Average MSE (all $r$) & Average MSE ($r>500m$) \\
             \hline
             MSE & -19.48dB & -23.4978dB\\
             \hline
             MSLE & -18.6718dB & -30.6704dB \\
             \hline
             WMSE $w=r$ & -19.36dB & -24.3503dB \\
             \hline
             WMSE $w= 1/(p_{\rm meas}+\epsilon)$ & -18.8631dB & -24dB \\
             \hline
        \end{tabularx}        
    \end{table}
    
\begin{figure}
        \centering
        \includegraphics[width=0.4\textwidth]{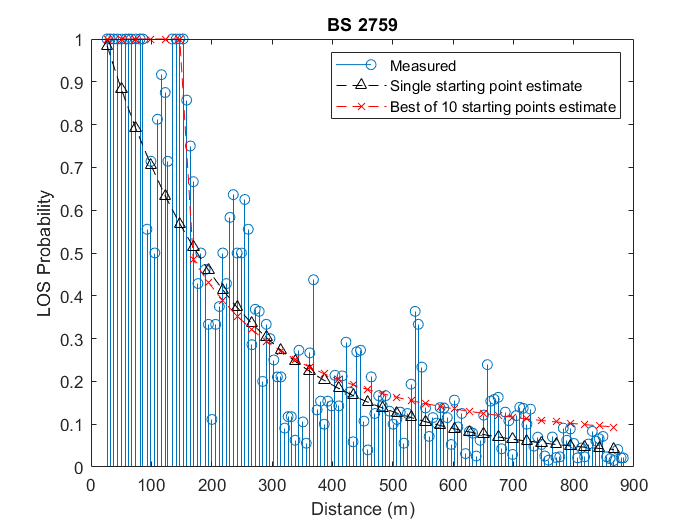}
        \caption{Comparison of the multi-starting points method (red cross) with the single starting point method (black triangles)}
        \label{fig:min10}
    \end{figure}

    % \item Show example comparison figure \textcolor{blue}{(0.25 column)} \\
    A comparison between the single starting point gradient descent and 10 starting points gradient descent is shown in \figurename \ref{fig:min10}. It can be seen that using the single starting point method provides an under-estimation of the cutoff distance which in turn drastically changes the values of the other parameters. Specifically, the single starting point optimal parameters are $(U^*=23.1,\:W^*=204.8,\: F^*=1)$, where we use $(.)^*$ notation to denote optimal parameters, and the ones using the multi-starting point method are $(U^*=158.5,\:W^*=108.2,\: F^*=0.51)$ which is a better fit both visually and numerically (-18.7dB \gls{MSE} for multi-starting point compared to -17.8dB \gls{MSE} for single starting point, about 23\% better).
    
    % \item BSs with not enough number of points, or irregular environment: Outliers and removal based on NSSE threshold \textcolor{blue}{(0.5 columns)} \\
    \subsection{Outlier detection}
\begin{figure}
        \centering
        \includegraphics[width=0.4\textwidth]{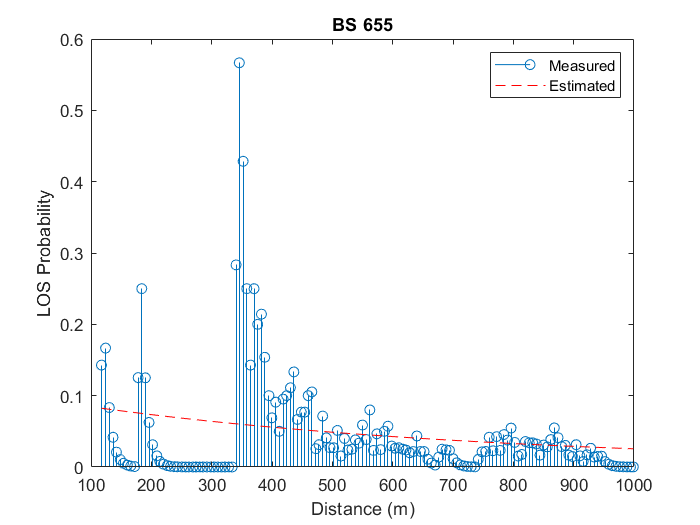}
        \caption{Example LOS probability of an outlier BS and its estimated LOS curve}
        \label{fig:outliers}
    \end{figure}

    There are some BSs that have an unusual distribution of LOS points due to the nature of the deployment and the environment, which leads to empirical LOS data that do not follow the standard exponential decay formulation. Such cases are labeled as outliers and are removed from subsequent processing. We use the \gls{NSSE} as the outlier detection criteria:
    \begin{align}
        \mathrm{NSSE} = \frac{\sum_{i=1}^{N} (p_{\mathrm{meas}}(r_i) - p_{\mathrm{LOS}}(r_i; U^*,W^*,F^*))^2 }{\sum_{i=1}^{N} p_{\mathrm{meas}}(r_i)^2 }
    \end{align}
    The idea is based on normalized residual error after fitting: if the \gls{NSSE} is greater than an empirically set threshold, the LOS model is not capturing the variation of the empirical data, therefore the BS is an outlier. The threshold is equal to 0.2 based on iteratively checking the results and adjusting the threshold. An example for such outliers is shown in \figurename \ref{fig:outliers}.
    
    % \item Figure of outlier \textcolor{blue}{(0.25 columns)} \\
        
    % \item Average model fitting: Same procedure, but combining all towers that have valid data together \textcolor{blue}{(0.25 columns)} \\
    The same fitting procedure explained above is used for fitting the average models per environment, except that we combine the data of all BSs in each environment class to create an average BS before binning. The optimal parameters of the average models are denoted as $U^*_{\mathrm{avg}}, W^*_{\mathrm{avg}}$ and $F^*_{\mathrm{avg}}$.
    
% \end{itemize}    

\section{Results} \label{sec:Results}
% \textcolor{red}{(around 7 columns)}
    %(Do we need a subsection per environment to describe what the average LOS probability looks like in each, and how it relates to the environment? Or do we include that in our analysis?)
    % \begin{itemize}
    % \item State limitations again: not taking vegetation into account (considered as LOS), discretization of streets with 10m raster (could have points closer to walls where LOS could change) \textcolor{blue}{(0.25 columns)} \\
    We now present the results for the LOS probability fits both for a ''typical" (average) cell and on a cell-by-cell basis. 
    We state again that the below results do not take into account vegetation and small cluttering objects, or different street widths that can lead to a different LOS condition depending on where a user stands on the street.
    % \item state how many towers are left after all the filtering \\
    The number of towers left after filtering using the 90\% height-to-footprint ratio, and NSSE-based outliers is 13,253, which is still far more than any previous work. We opted to filter based on the 90\% threshold to ensure correctness and reliability of results instead of having more cells but with unreliable GIS data. Note that after filtering with the height-to-footprint ratio,  only 2\% of the remaining towers are classified as outliers.%, but possibly unreliable, BSs.
    \subsection{Environmental Average models} 
\begin{figure}
                \centering
                \includegraphics[width=0.4\textwidth]{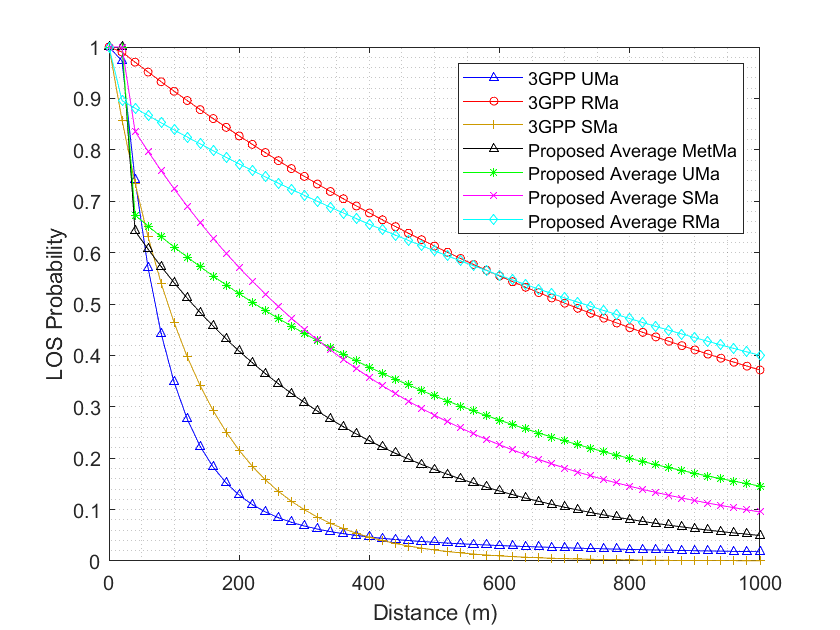}
                \caption{Overlay plot illustrating the models for each environment compared to 3GPP models.}
                \label{fig:models-plots}
            \end{figure}

    % \textcolor{blue}{(1 col)}
        % \begin{itemize}
            % \item Best parameter fits for average models and comparison with optimal parameters if we were to use the $d_1/d_2$ and 3GPP models, resulting SSE \\
            We present the results of the average models for each environment and compare them to 3GPP (using 3GPP-UMa with $d_1=18m$, $d_2=63m$ for the \gls{UMa} and \gls{MetMa} environments, and 3GPP-RMa for the rest) and also compare to the $d_1/d_2$ model when $d_1$ and $d_2$ are optimized. The comparisons are in terms of MSE in Table \ref{table:avg-results}. 
            %%%%%%%%% NEW: %%%%%%%%%%%%%%%%%%
            Note that these models are for all cells including outliers in each environment, to show the raw result for all deployments. If we filter out the outliers, we get slightly different parameters, and the discontinuity gets smoothed out on average. Namely we get: $(5, 487,0.97)$ for the urban environment, $(5, 487,0.97)$ for the metropolitan, $(34,400,0.95)$ for the suburban, and $(44,1098.6,0.98)$ for the rural environment. This implies outliers need special processing, which will be part of a future work since they are a result of deployments that result in LOS curves that significantly deviate not only from the 3GPP base model and its extensions, but from any monotonously decreasing function. Furthermore, these parameters of the average models cannot be obtained by computing the means of the distributions of the $(U,W,F)$ parameters. The reason is that creating an average environment leads to a new \gls{LOS} probability curve that is obtained by averaging the {\em curves} generated using the $(U,W,F)$ parameters' distributions.
            %%%%%%%%%%%% END NEW %%%%%%%%%%%%%%%
            Also note that for all cases our models achieve lower MSE than the 3GPP models.
            % \textcolor{blue}{(0.25 - 0.5 columns)}
            % \item Discussion and comparison of models to 3GPP \\
            The scale parameter for each average model is less than 1, which shows the necessity of modeling it, however it gets closer to 1 when going from more urbanized, dense environments to less urbanized ones. This phenomenon is expected because less urbanized environments have shorter buildings, making the LOS conditions more favorable in general. Also, the cutoff distance and decay parameter decrease as we go from less urbanized to more urbanized environments due to the same reason, showing less favorable LOS conditions as urbanization increases. A plot comparing the average models to the 3GPP-UMa, 3GPP-SMa and 3GPP-RMa models is shown in \figurename~\ref{fig:models-plots}. 
            % \item Figure of average fit for all environments and  (?) \textcolor{blue}{(0.25 columns)} \\
            % \item Estimated parameters and SSE
            % \item Table: Average fit results per environment along with optimal 3GPP parameters per environment (with SSE for each) \textcolor{blue}{(0.25 column)} \\                    
            \begin{table}[]
                \centering
                \caption{Per environment average models and comparisons with 3GPP fits}
                \label{table:avg-results}
                \begin{tabularx}{0.5\textwidth}{|>{\centering\arraybackslash}X
                |>{\centering\arraybackslash}X
                |>{\centering\arraybackslash}X
                |>{\centering\arraybackslash}X
                |>{\centering\arraybackslash}X|}
                     \hline
                     Environment & MetMa & UMa & SMa & RMa \\
                     \hline
                     \textbf{($U_{\rm avg}^*$, $W_{\rm avg}^*$, $F_{\rm avg}^*)$} & (22.1, 339.5, 0.6756) & (21.9, 607.4, 0.6929) & (33, 400, 0.85) & (9.9, 1209.6, 0.9031)\\
                     \hline
                     \textbf{$(d_1,d_2)$} & (2, 240.6) & (10, 396.2) & (10.2, 368.2) & (10, 1169.2)\\
                     \hline
                     MSE & -27.0235dB & -28.9919dB & -24.5dB & -23.9149dB \\
                     \hline
                     $\mathrm{MSE}_{d_1/d_2}$ & -22.2032dB & -21.78dB & -22.9224dB & -21.0957dB\\
                     \hline 
                     $\mathrm{MSE}_{\rm 3GPP}$ & -10.9dB & -15.5dB & -10.37dB& -22dB\\  
                     \hline
                \end{tabularx}                
            \end{table}
            
        % \end{itemize}
\subsection{Cell-by-cell models: Motivation and examples}
            % \item Average SSE  \textcolor{blue}{(less than 0.25 columns)} \\
% While the MSE of the average model fit is small (-20 dB), 
While the MSE of the per-environment average models is small for our model (below -20dB), we stress that this is the deviation between the {\em average} empirical data against the average model. However, when considering the error of each individual cell compared to the overall average model across all environments, the result is much larger, namely -11.7dB. 
Conversely, when we fit on a cell-by-cell basis, then the MSE (averaged over the cells) is -21.54dB, indicating a good fit using our method. 
This clearly motivates modeling the LOS probability on a cell-by-cell basis. 

\begin{figure}
    \centering
      \subfloat[LOS probability fit for BS~1519\label{fig:tower1519}]{
        \includegraphics[width=0.4\textwidth]{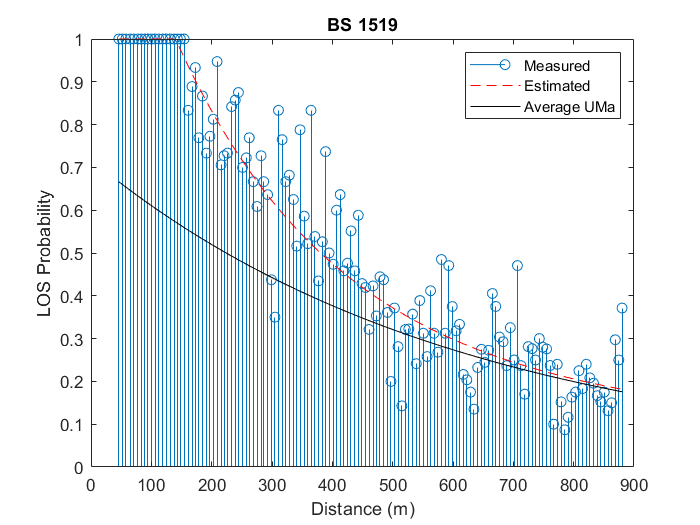}
      }\hfill
      \subfloat[LOS probability fit for BS~1832\label{fig:tower1832}]{
        \includegraphics[width=0.4\textwidth]{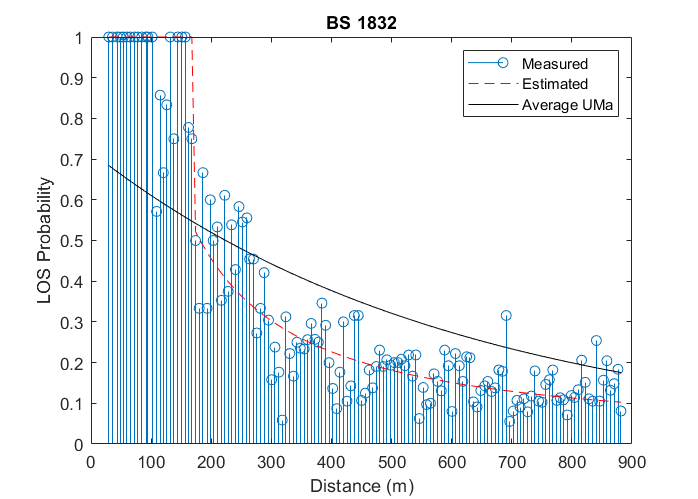}
      }
    \caption{Sample cell-by-cell LOS probability fitting examples (a) BS 1519 and (b) BS 1832 with comparison to average UMa model}
    \label{fig:tower-fits}
\end{figure}

Fig. \ref{fig:tower-fits} shows an example of a per-cell fit, where we can clearly see a discontinuity that needs to be modeled to ensure a good fit.
% \textcolor{blue}{(1 column)} \\
% \begin{itemize}         
% \item Figure: Graphical results of 2 towers \textcolor{blue}{(0.5 columns)} \\
% \item Mention error modeling \textcolor{blue}{(less than 0.25 clumns)} \\
A further refinement of the model can be achieved by detailed characterization of the difference between the empirical probability and the fitted probability, which will be addressed in future work.

\begin{figure}[h]
     \centering
      \subfloat[UE height = 0 m\label{fig:scatt_h=0m}]{
        \includegraphics[width=0.4\textwidth]{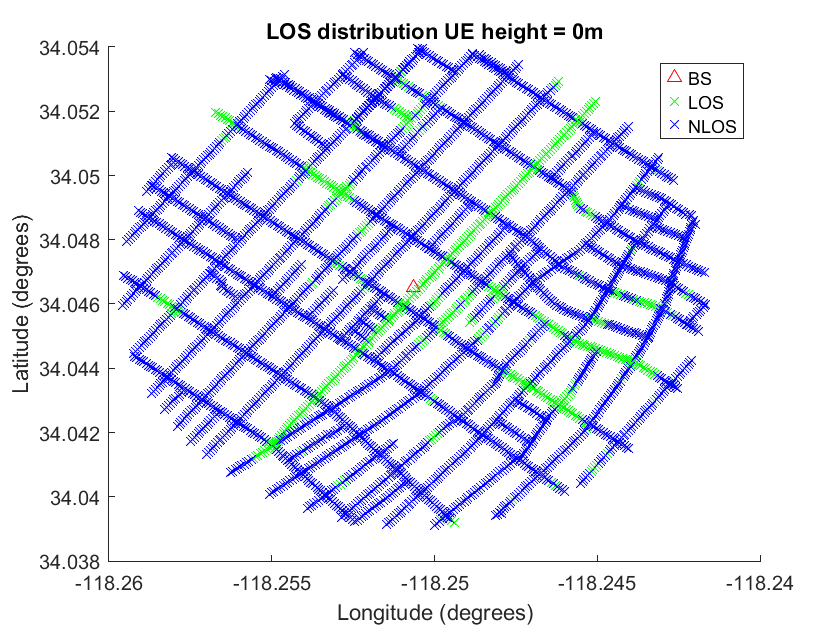}
      }\hfill
      % Rename file to avoid comma (recommended): scatter_ue_height=1,5m.png
      \subfloat[UE height = 1.5 m\label{fig:scatt_h=1.5m}]{
        \includegraphics[width=0.4\textwidth]{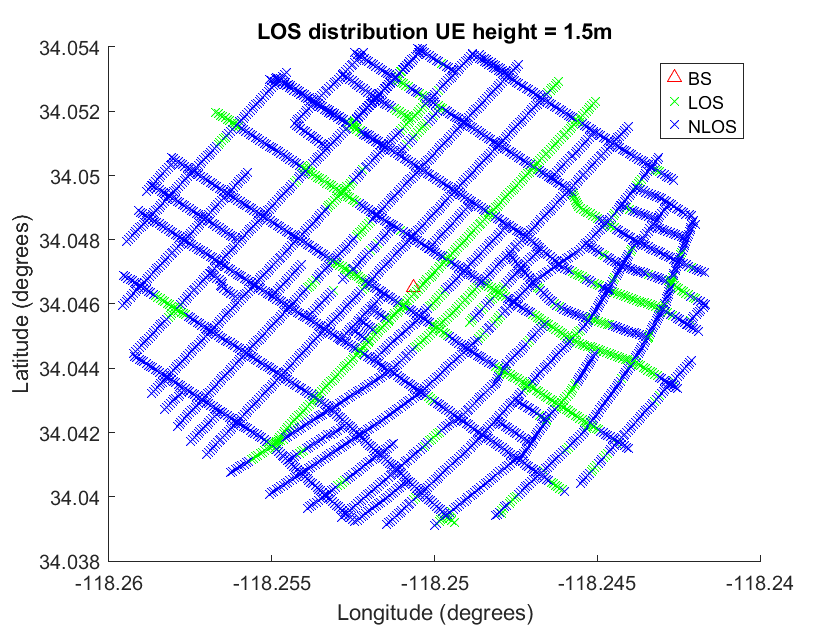}
      }
     \caption{Comparison of LOS distribution for UE height = 0m and 1.5m}
     \label{fig:scatter_comparison_heights}
\end{figure}

\begin{figure}[h]
     \centering
      \subfloat[UE height = 0 m\label{fig:plos_h=0m}]{
        \includegraphics[width=0.4\textwidth]{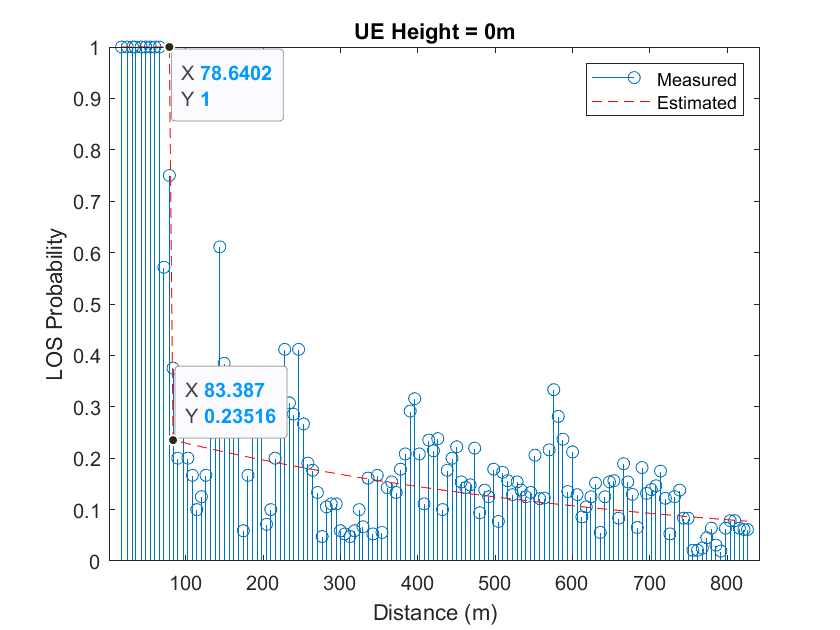}
      }\hfill
      % Rename file to avoid comma: plos_ue_height=1_5.png
      \subfloat[UE height = 1.5 m\label{fig:plos_h=1.5m}]{
        \includegraphics[width=0.4\textwidth]{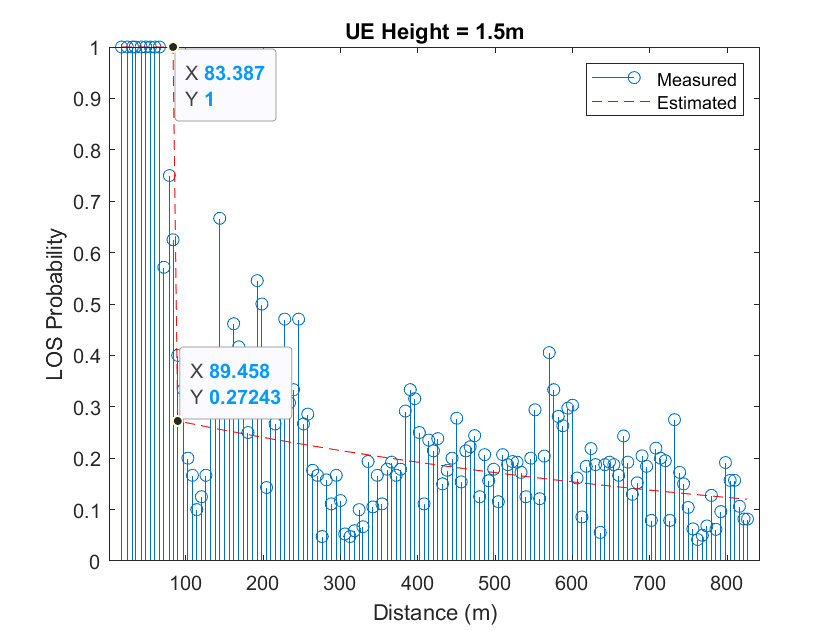}
      }
     \caption{Comparison of LOS probability curves for UE height = 0m and 1.5m}
     \label{fig:plos_comparison_heights}
\end{figure}

\textcolor{black}{Regarding the higher \gls{UE} height as discussed in Sec. II, we found that it does change the \gls{LOS} probability but not to a significant extent. An example comparison of how a height of 1.5m affects the \gls{LOS} status of street points and consequently the \gls{LOS} probability is shown in Fig. \ref{fig:scatter_comparison_heights} and \ref{fig:plos_comparison_heights} for a \gls{BS} in an urban environment in downtown Los Angeles. The number of \gls{LOS} points naturally increases when the street point heights increases, which leads to an increase in the cutoff distance $U$ by only one distance bin (5m) from 78m to 83m. The scaling parameter $F$ increases from 0.2352 to 0.2724, meaning the discontinuity between regions decrease. Although the \gls{LOS} probability is slightly higher for distances greater than $U$, the same decay parameter $W = 838$ is still a good fit, providing a fitting MSE of -20.7dB for the case of UE heights = 0m, and -18.9dB for UE heights = 1.5m. }

        % \end{itemize}
        
\subsection{Environmental distribution fitting}
            A key part of our contribution is modeling the LOS model parameters as random variables within each environment. We fit the distribution of the parameters using the process outlined below. We also study the correlations between parameters in each environment.
            
            % \item Explanation of possible PDFs to use due to restrictions on each parameter \textcolor{blue}{(0.25 columns)} \\
            Note that there is a limited set of candidate distribution families for the cdf of each parameter. This is partly motivated by the need to use well-known distributions with a small number of parameters, and partly because of the restrictions on their values. Namely, $U$ and $W$ cannot be negative, and $F$ can only be between 0 and 1. For those reasons, our candidate distributions for $U$ and $W$ are: gamma, exponential and generalized extreme value (GEV). For $F$, we are limited to the uniform and beta distributions since they are defined on the interval $[0,1]$.
            The gamma, exponential, GEV and beta distributions are respectively defined as:
            \begin{align}
                f_X(x) &=  \frac{1}{\Gamma(k) \theta^k}x^{k-1}e^{-\frac{x}{\theta}} \\
                f_X(x) &= \frac{1}{\theta}e^{-\frac{x}{\theta}}\\
                f_X(x) &= \frac{1}{\sigma}e^{-(1+ k \frac{(x-\mu)}{\sigma})^{-\frac{1}{k}}}(1+ k \frac{(x-\mu)}{\sigma})^{-1-\frac{1}{k}}\\
                f_X(x) &= \frac{x^{\alpha - 1}(1-x)^{\beta-1}}{ B(\alpha,\beta)}
            \end{align}
            where $X$ is the parameter under consideration, $\Gamma(.)$ is the gamma function, and $B(.,.)$ is the beta function \cite{abramowitz1968handbook}.
            
            % \item AIC based distribution estimation explanation and Delta AIC thresholds \textcolor{blue}{(0.5 column)} \\
            We fit the distributions using \gls{ML} estimation of the parameters $\bm{\theta}$ for each candidate distribution family $j$ (denoted as $\hat{\bm{\theta}}_j$) and we use the \gls{AIC}, corrected for small sample size (\gls{AICc}) \cite{aicc}, to choose the best candidate family. We choose the candidate distribution family that gives the lowest \gls{AICc} score $\mathrm{AICc}_{\rm min}$. We further compute the $\Delta \mathrm{AICc}_j = \mathrm{AICc}_j - \mathrm{AICc}_\mathrm{min}$. This is done to check if there are other plausible distributions that could explain the data well and have lower complexity in terms of number of parameters of the distribution. Some researches, e.g. \cite{aic-book}, use a threshold on the $\Delta \mathrm{AIC}$ (or $\Delta \mathrm{AICc}$) to select competing candidate distributions: a $\Delta \mathrm{AIC}$ less than 2 means that the considered competing distribution is just as likely to be the real distribution, a $\Delta \mathrm{AIC}$ in the range of 2 to 7 also means that the competing distribution is valid but it is less plausible that it is the real distribution.
            % The candidate family with the lowest $\Delta \mathrm{AICc}_j = \mathrm{AICc}_j - \mathrm{AICc}_\mathrm{min}$ is chosen as the best fit, where $\mathrm{AICc}_\mathrm{min}$ is the lowest \gls{AICc} score of the candidate distribution after \gls{ML} estimation, and $\mathrm{AICc}_j$ is the \gls{AICc} score of the $j$-th candidate distribution. Note that some researches, e.g. \cite{aic-book}, use a threshold on the $\Delta \mathrm{AIC}$ (or $\Delta \mathrm{AICc}$) to select competing candidate distributions: a $\Delta \mathrm{AIC}$ less than 2 means that the considered competing distribution is just as likely to be the real distribution, a $\Delta \mathrm{AIC}$ in the range of 2 to 7 also means that the competing distribution is valid but it is less plausible that it is the real distribution.

            % \item Steps and equations for fitting \textcolor{blue}{(0.5 column)}
            
            The steps for the fitting procedure are outlined below:
            \begin{itemize}
                \item For each candidate distribution family indexed by $j$, find the \gls{ML} parameter estimates:
                \begin{align}
                    \hat{\bm{\eta}}_j = \argmax_{\bm{\eta}_j} \ln (f_{\mathrm{th},j}(x,\bm{\eta}_j))
                \end{align}
                where $f_{\mathrm{th},j}(x,\bm{\eta}_j)$ is the theoretical distribution of parameter $x$ assuming distribution family $j$ with parameters $\bm{\eta}_j$.
                \item For each of the candidate distributions $j$ compute the \gls{AICc} scores using the \gls{ML} parameters.
                \begin{align}
                    \mathrm{AIC}_j &= 2k - 2\ln(f_{th,j}(x,\hat{\bm{\eta}}_j))\\
                    \mathrm{AICc}_j &= \mathrm{AIC} + \frac{2k^2 +2k}{n-k-1}
                \end{align}
                where $k$ is the number of parameters that describe the distribution (length of the vector $\bm{\eta}_j$) and $n$ is the number of observations.
                \item Compute $\Delta \mathrm{AICc}$ score for each candidate distribution $j$
                \begin{align}
                    \Delta \mathrm{AICc}_j = \mathrm{AICc}_j - \mathrm{AICc}_\mathrm{min}
                \end{align}
                \item If there exist candidate distributions $j$ such that $0 \leq \Delta \mathrm{AICc}_j \leq 7$, choose the one that has the lowest number of parameters (lowest complexity) as the best fit distribution. Otherwise, select the distribution $j$ that corresponds to the lowest $\mathrm{AICc}$ score.
            \end{itemize}

            % \item Table showing distribution of each parameter in each environment. comment on Met fitting with lower HFI threshold \textcolor{blue}{(0.5 column)} \\
            The results of the fitting for each parameter are shown in Table \ref{table:env-results} as well as the percentage of \gls{BS}s for each environment after filtering. Note that the filtered towers for the metropolitan case using the 90\% threshold of height-to-footprint ratio are mostly located in Manhattan. If we use a less stringent constraint, e.g. 80\% threshold, for the sake of diverse locations, we obtain 1.39\% of the total towers, located in Manhattan, Los Angeles, and San Francisco. The resulting parameter distributions in that case for $U,\: W, \: F$ are respectively: Gamma($k=0.4493, \theta = 150.77$), Gamma($k=0.3339,\theta = 1398$), Beta($\alpha = 0.7926,\beta = 0.4191$).
            Note that for the $U$ distribution in the rural case, there is a higher probability that $U$ is in the range close to 1000m than the rest of the 0m to 1000m range. This might not necessarily mean that $U=1000$m exactly, because we cut off our radius of interest at 1000m, but there could be unobstructed LOS for a larger distance due to the nature of the rural environment where minimal obstructions are present.
            %%%%%%%%% The values for the distributions need to change according to new results of lin Prob - log dist fitting
            \begin{table}[]
                \centering
                \caption{Estimated distributions for each parameter in each environment}
                \label{table:env-results}
                \begin{tabularx}{0.5\textwidth}{|>{\centering\arraybackslash}X
                |>{\centering\arraybackslash}X
                |>{\centering\arraybackslash}X
                |>{\centering\arraybackslash}X
                |>{\centering\arraybackslash}X|}
                     \hline
                     \textbf{Environ- ment} & MetMa & UMa & SMa & RMa \\
                     \hline
                     \textbf{$U$ distribution} & Gamma($k=0.1124, \theta = 752.3$) &  Gamma($k=0.2352,\theta=531.29$) & Gamma($k = 0.2039,\theta = 501.42$)& Gamma($k= 0.2932, \theta = 1126$) \\ %Gamma($k=0.2731,\theta=317.1502$)
                     \hline
                     \textbf{$W$ distribution} & Gamma($k=0.4223,\theta = 2242.9$) & Gamma($k=0.7759, \theta = 849.43$) & Gamma($k=0.7556, \theta = 687.66$) & Gamma($k = 0.4679,\theta = 2937$)\\
                     \hline
                     \textbf{$F$ distribution} & Beta($\alpha =0.5276,\beta = 0.1691$) & Beta($\alpha = 0.4266,\beta=0.1204$) & Beta($\alpha = 0.3962, \beta = 0.1035$) & Beta($\alpha = 0.4124,\beta =0.1951$)\\
                     \hline
                     \textbf{Percentage of BSs} & 0.86\% & 15.04\% & 60.7\% & 23.3\% \\
                     \hline
                \end{tabularx}                
            \end{table}
            % \item Interpolation of building heights comment \\
            Also note that interpolating the missing building heights using the surrounding buildings only results in about 1.6\% change in the LOS labels of street points on average.
        % \end{itemize}
    % \item Tables showing correlations between all parameters in all environments \textcolor{blue}{(0.5 column)} \\
    To generate the LOS parameters for simulation purposes, the correlation of the parameters needs to be studied. Table \ref{tab:corr-table} shows the linear correlation coefficients between each parameter pair in all environments. The correlation between all pairs of parameters is relatively weak, with the most correlated pair being $(U,F)$ with a negative correlation. This is expected since the larger the cutoff distance, the more favorable the LOS conditions are, hence the discontinuity between the constant and decaying regions will be smaller.
    \begin{table}
        \centering
        \caption{Pairwise correlations between LOS parameters in each environment}
        \label{tab:corr-table}
        \begin{tabular}{|c|c|c|c|}
            \hline
             Environment & $(U,W)$ & $(U,F)$ & $(W,F)$\\
             \hline
             RMa & 0.0039 & -0.3094 & 0.0046\\
             \hline
             SMa & -0.0079 & -0.2147 & -0.174\\
             \hline
              UMa & -0.0362 & -0.1615 & -0.1164\\
             \hline
             MetMa & 0.069 & -0.4423 & -0.3306\\
             \hline
        \end{tabular}        
    \end{table}

Since the parameters are correlated, the procedure to generate the correlated $(U,W,F)$ triplet is inspired by the inverse \gls{CDF} transform and \cite{copula} and is the following:
            \begin{enumerate}
                \item Define the correlation matrix for the chosen environment based on Table \ref{tab:corr-table}
                \begin{align}
                    \mathbf{R} = \begin{bmatrix}
                        1 & \rho_{UW} & \rho_{UF} \\
                        \rho_{UW} & 1 & \rho_{WF} \\
                        \rho_{UF} & \rho_{WF} & 1
                        \end{bmatrix}
                        \label{eq:corr_matrix}
                \end{align}
                where $\rho_{UW}$, $\rho_{UF}$ and $\rho_{WF}$ are the correlation coefficients of the chosen environment from Table \ref{tab:corr-table}.
                \item Perform Cholesky decomposition of $\mathbf{R}$ such that $\mathbf{R} = \mathbf{L}\mathbf{L}^T$
                where $\mathbf{L}$ is a lower triangular matrix.
                \item Generate independent identically distributed standard normal random variables $Z_1$, $Z_2$, $Z_3$.
                \item Generate correlated normal random variables $X_1$, $X_2$, $X_3$ using
                $\begin{bmatrix} X_1, X_2, X_3 \end{bmatrix}^T = \mathbf{L} \begin{bmatrix} Z_1, Z_2 Z_3 \end{bmatrix}^T$
                \item Apply the inverse \gls{CDF} of the corresponding distributions in Table \ref{table:env-results} to each of $X_1$, $X_2$, $X_3$ to obtain $U$, $W$ and $F$ respectively, using $U = F_{U}^{-1}(F_{X_1}(X_1))$, $W = F_{W}^{-1}(F_{X_2}(X_2))$, $F = F_{F}^{-1}(F_{X_3}(X_3))$, where $F_{X_1}(.),\: F_{X_2}(.), \: F_{X_3}(.)$ are the standard normal CDFs, $F_{\eta}^{-1}(.)$ is the inverse \gls{CDF} of the corresponding distribution for parameter $\eta$.
            \end{enumerate}
            A key assumption is the invertibility of the \gls{CDF} of each parameter, which is guaranteed for the estimated distributions.\
                           
    % \item Comment on complexity of chosen distribution and peaks (Note: I need to check if the peaks still exist with the latest data, my guess is no since it's OSM data that's more accurate) \\
    
    % Note that the resulting best fit distributions have a slightly high complexity in terms of number of parameters needed to describe them. The reason is that our results are purely based on the data and statistical methods without making any a-priori assumptions on how the parameters should behave. 
    % We observed a slightly higher probability in the $U$ distribution in the range of 900m to 1000m in the rural case, which can be explained by the low building heights in such environments that allows for a high LOS probability, hence a larger region where the probability is 1. This peak is only about 6\% of the data, and could be modeled as an additional Dirac delta impulse to the $U$ distribution, however that would increase the complexity of resulting distribution and would make it harder to generate values for system level simulations, so we choose to not model the peak due to its small relative to the rest of the data.  

\subsection{Outage Simulation}

\begin{figure}
        \centering
        \includegraphics[width=0.5\linewidth]{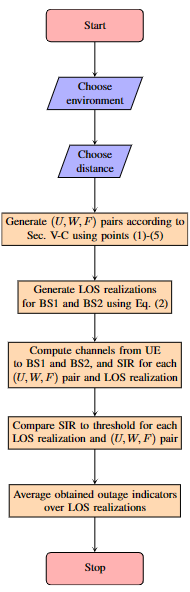}
        \caption{Simulation procedure to study outages for a particular environment.}
        \label{fig:flowchart}
    \end{figure}

        % Flowchart similar to the one in the conference paper as to how to generate a model, with all its parameters, for a certain scenario. \textcolor{blue}{(0.5 columns)}
    Fig. \ref{fig:flowchart} provides a flowchart for the simulation procedure. After selecting the environment, we determine the correlated variables $X_1$, $X_2$, $X_3$ according to the procedure of (\ref{eq:corr_matrix}) and the subsequent bullet items 2-4. From that the realizations of $U$, $W$, and $F$ using the \gls{CDF}s from Table \ref{table:env-results} are created according to item 5). For system simulations, this is done independently for each cell. With this, the LOS probability in each cell can be computed from (\ref{equ:plosmodel}).  For each UE location, the LOS status to the different BSs is then determined probabilistically, according the various computed LOS probabilities. According to this status, the pathloss model is chosen. 
    % \end{itemize}
    
    % \textcolor{blue}{(2 columns)} \\
    We use a link-level Monte Carlo simulation to compare the predicted outage probability as computed with our model relative to that obtained from the 3GPP-UMa model in urban scenarios. Since the 3GPP UMa model is not specified for any \gls{BS} height greater than 25m and because our model is height independent, we use a BS height of 25m for the computation of the 3D distances that are required in the pathloss model. 
    %our \gls{UMa} model with a BS height of 25m to compare with to the 3GPP UMa model.
    %For a fair comparison with 3GPP, we use our urban model with a BS height of 25m to compare to the 3GPP UMa model. 
    Note, that we compare here our UMa model with that of 3GPP; since 3GPP does not separately define a metropolitan environment, the errors induced by using the 3GPP-UMa model for this case would be larger.
        % \begin{itemize}
            % \item Scenario explanation: number of BSs and distances. What models are used, number of realizations \textcolor{blue}{(0.5 - 0.75 columns)} \\
           
            The scenario we consider is where one \textcolor{black}{pedestrian} \gls{UE} \textcolor{black}{(at ground level)} is connected to one serving \gls{BS} (BS1) and there is one interfering \gls{BS} (BS2). The distance from the \gls{UE} to BS1 and BS2, respectively $d_{\mathrm{BS1}}$ and $d_{\mathrm{BS2}}$, are varied to study the outage distributions for different distances ranging from close to the serving \gls{BS}, to equally distant from the serving and interfering BS, simulating a cell-edge case. \textcolor{black}{Regarding the impact of higher \gls{UE} height, we mentioned above that a height of 1.5m does not have a significant impact on the \gls{LOS} parameters, hence we expect similar outcomes and insights as will be presented here for the case of 1.5m UE height}. We assume the cells have a radius of 500m and thus $d_{\rm BS2}$ is given by $d_{\rm BS2} = 1000 - d_{\rm BS1}$. Due to the random nature of the parameters of our model, we sample 1000 different pairs of realizations of the ($U,W,F$) triplet (for BS1 and BS2) to generate 1000 LOS realizations for each pair and calculate the SIR for each of the $10^6$ realizations.
            %%%%% Correlation of parameters generation %%%%%%
            \
            %%%%
            The pathloss and shadowing models are the same as the 3GPP specifications in \cite{38901}: the shadowing is log-normally distributed with standard deviation in the \gls{LOS} case $\sigma_{\mathrm{LOS}} = 4\mathrm{dB}$ and in the \gls{NLOS} case $\sigma_{\mathrm{NLOS} } = 6 \mathrm{dB}$.
            % \item Small scale fading average \\
            We assume that the small scale fading is averaged out by spatial and/or frequency diversity schemes.
            % \item Choice of operating frequency, MCS number\\
            We use an operating frequency of 740MHz, corresponding to band n12 in 3GPP New Radio (NR) \cite[Chapter32]{molisch2023wireless}. The SIR threshold we set for outage calculations is 0.399dB, the minimum operating SIR for the modulation and coding scheme (MCS) number 5 according to the NR standard \cite{mcs-nr}. As per the procedure described at the beginning of this subsection, in each run of the Monte Carlo simulation, the \gls{LOS} probabilities of each link are determined by sampling a random number from the standard uniform distribution (between 0 and 1) and comparing it to the $p_{\mathrm{LOS}}$ at the UE location, if the sampled value is less than $p_{\mathrm{LOS}}$ the link is \gls{LOS}, and \gls{NLOS} otherwise.
            
            % \item SIR distribution results comparisons, and coverage probability comparisons \textcolor{blue}{(0.75 column)} \\
            
\begin{figure}
                \centering
                \includegraphics[width=0.45\textwidth]{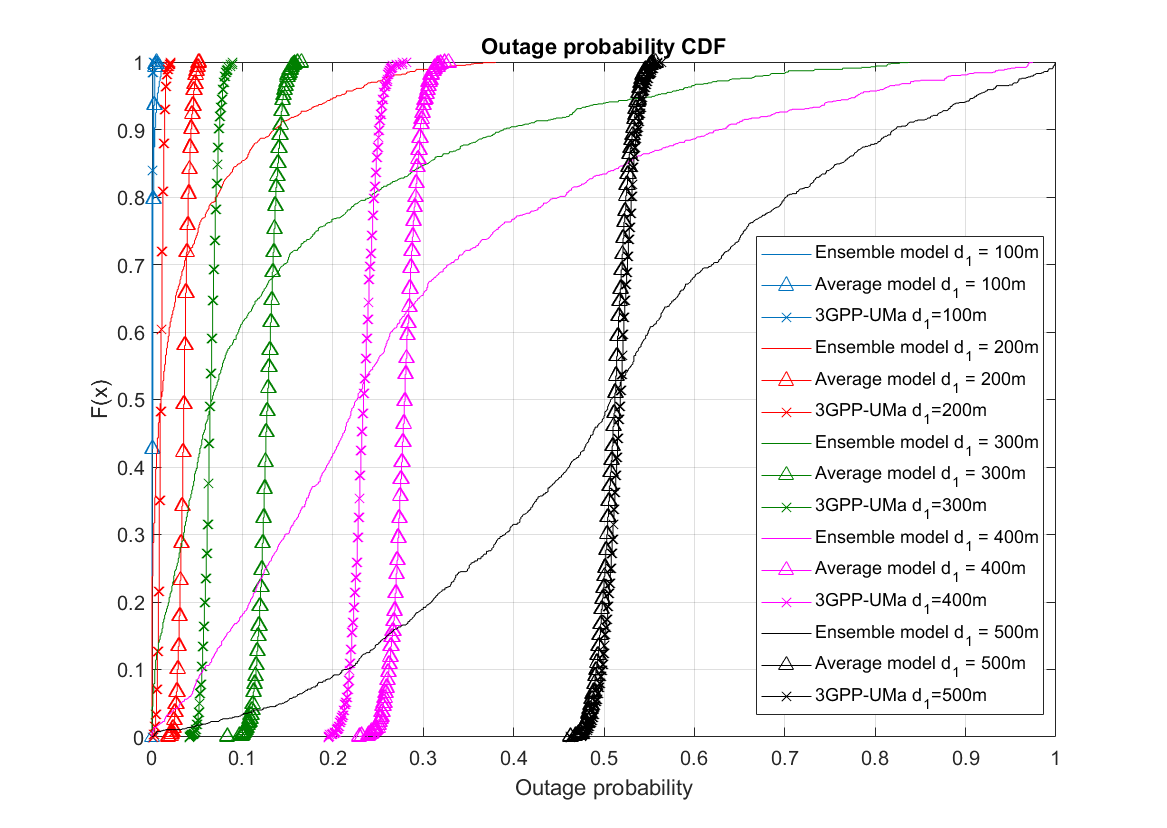}
                \caption{Outage probability CDFs comparison for different UE distances under different LOS models.}
                \label{fig:outage-cdf}
            \end{figure}

            We define the outage probability at a certain UE distance as the mean number of times the SIR crosses below the threshold, averaged over the number of LOS realizations at that particular distance.
            We study the distribution of the outage probability over all cells (i.e., over all $(U,W,F)$ pairs) for different UE distances under the ensemble model (generating $(U,W,F)$ pairs according to the procedure in Sec. V-C points (1) to (5) and aggregating the outage values), average UMa model (Table \ref{table:avg-results}) and 3GPP-UMa model.
            The resulting CDFs are shown in Fig. \ref{fig:outage-cdf}.
            On average, the outage probability increases with distance for all models as expected, with the average model and 3GPP-UMa having similar outage predictions, and no outages for UE distance $d_{\rm BS1}=100m$ to BS1. Both of those models have a low variance in the outages predicted for each cell due to the deterministic nature of their parameters. For the ensemble models, we also have an increase in outage probability on average with distance, as well as an increase in variance of the outage values due to the different combinations of $(U,W,F)$. This larger variance becomes more important as UE distance to the serving BS increases, since there is a larger probability that certain combinations of cells (pair of $(U,W,F)$ triplets) will result in high outage probability, as well as a higher probability that certain combinations of cells result in outage values lower than the ones predicted by 3GPP-UMa and the average model.

            %%%%%%%%%%%%% NEW %%%%%%%%%%%%%%%%
            Additional simulations were performed for the case of multiple interfering \glspl{BS} and including beamforming from all \glspl{BS}. Apart from a shift in the average signal-to-interference-and-noise-ratio (SINR) due to the beamforming and hence the average outage probability, the qualitative results are the same as the ones we present in \figurename~\ref{fig:outage-cdf}.            
            %%%%%%%%%%%%% END NEW %%%%%%%%%%%%%

            Note that this is a worst case analysis, since in practice, the parameters $(U,W,F)$ might be correlated between neighboring cells, thus reducing the likelihood of having cells with drastically different $(U,W,F)$ parameters next to each other. which in turn  
%            in the same geographical vicinity such that we have very high (or low) outage probability for a \gls{UE} at a certain distance from its serving \gls{BS} might not be very high (and 
could lead to smaller variance in the outage CDFs. This phenomenon remains to be studied. 
            
            % Our proposed models predict a larger variance of the SIR as seen in \figurename \ref{fig:sir-cdf}, meaning that the channel conditions could be more hostile or more beneficial, then one would anticipate from the 3GPP model. We show the \gls{CDF}s for the statistical ensemble UMa model (sampling $(U,W,F)$ triplets and calculating the SIR), the average UMa model (using parameters in Table \ref{table:env-results}) and 3GPP.
            % \item Figure of SIR CDF for both cases, 3GPP and our model. \textcolor{blue}{(0.25 column)} \\

\section{Conclusion} \label{sec:Conclusion}
% \textcolor{red}{(0.5 columns)}
        % \begin{itemize}
            % \item Summary of the purpose of the work \\
            In this paper, we provide a complete framework to generate realistic LOS probability models. We used this framework to generate both cell-specific and environmental-average models based on a nation-wide geospatial database of BSs in the continental US. 
            % \item Re-stating what differentiates our paper from others \\
            Our work differentiates itself from previous works by the large number of real-world BS positions used, and by expanding the modeling structure of the 3GPP $d_1/d_2$ model by introducing a scaling parameter describing jumps in the LOS probability. Most importantly, we propose that modeling needs to be done on a cell-by-cell basis, and thus the LOS model parameter should be described as random variables with a particular distribution.
            % \item Summary of results and stating the limitations again \\
            We found that 3GPP models under-estimate the LOS probability overall, and in particular at the cell-edge, due to certain assumptions, and limited measurement and/or simulation data used to create those models. Our model thus creates a more realistic assessment of outage probability. 
            
            The limitation of our current approach is the reliability of the underlying geospatial database, which as we explained, is crowd-sourced and incomplete in some cases, leading to the need for further filtering to keep the reliable data only. It is possible that a commercial database would provide improvements of the model reliability and accuracy. Such an improved model can be easily derived using the framework of our paper.
            Furthermore, while our results are based on a much larger database than before, they are still specific to one country, i.e., the US. Evaluations in other countries, in particular Asia, South America, and Africa, which had not been considered by 3GPP either, should be done with databases available there, and again can follow our framework.
            % \item Future work: Error Modeling \\
            Future work will include modeling the error between empirical probability and the model fit in order to further increase the model accuracy.
        % \end{itemize}

\bibliography{reference.bib}{}
\bibliographystyle{IEEEtran}
\end{document}